\documentclass[12pt, a4paper]{article}

\usepackage[
  margin=0.75in,
  headsep=10pt, 
]{geometry}

\usepackage{graphicx}
\usepackage{soul}
\usepackage{graphicx}
\usepackage{epstopdf}
\usepackage{epsfig}
\DeclareGraphicsExtensions{.pdf,.eps,.jpg,.png}
\usepackage{amsmath}
\usepackage[table]{xcolor}
\usepackage{subcaption}
\usepackage{soul}
\usepackage{makecell}
\usepackage{multirow}
\usepackage{breqn}
\usepackage{amssymb} 
\usepackage{booktabs}
\usepackage{dcolumn}
\usepackage{bm}
\usepackage[utf8]{inputenc}
\usepackage[T1]{fontenc}
\usepackage{mathptmx}
\usepackage{etoolbox}
\usepackage{hyperref}
\usepackage{subcaption}
\usepackage{ragged2e}
\usepackage[sort&compress]{natbib} 

\newcommand{\edt}[1]{{\color{black}#1}} 
\newcommand{\ahf}[1]{{\color{black}#1}} 
\newcommand{\fnl}[1]{{\color{black}#1}} 

\newcommand{\soptitle}{Characterizing the Role of Hind Flippers in Hydrodynamics of A Harbor Seal}

\usepackage{xcolor} 

\begin{document}


\begin{center}
\Large \bf{\soptitle}
\vspace{0.1in}
\end{center}

\begin{center}
{Amirhossein Fardi$^1$, Hamayun Farooq$^2$, Imran Akhtar$^3$, Arman Hemmati$^4$, and Muhammad Saif Ullah Khalid$^{1\star}$}\\
\vspace{0.1in}
\end{center}
\begin{center}
$^1$Nature-Inspired Engineering Research Lab (NIERL), Department of Mechanical \& Mechatronics Engineering, Lakehead University, Thunder Bay, ON P7B 5E1, Canada\\
$^2$Department of Mathematics, Emerson University, Multan, 60700, Pakistan\\
$^3$School of Interdisciplinary Engineering \& Sciences, National University of Sciences \& Technology, Islamabad 44000, Pakistan\\
$^4$Department of Mechanical Engineering, University of Alberta, Edmonton, AB T6G 1H9, Canada\\
\vspace{0.05in}
$^\star$\small{Corresponding Author, Email: mkhalid7@lakeheadu.ca}
\end{center}

\begin{abstract}

In this paper, we investigate the hydrodynamic characteristics of harbor seal locomotion, focusing on the role of hind flippers in thrust generation and wake dynamics. Through three-dimensional numerical simulations using an immersed boundary method at Reynolds number \fnl{of} $3000$, we analyze the impact of varying Strouhal number ($\mbox{St} = 0.2-0.35$) and propulsive wavelength ($\lambda^\ast = 1.0-1.2$) on swimming performance. Our findings reveal two distinct wake patterns: a single-row structure at lower Strouhal numbers ($\mbox{St} \leq 0.25$) and a double-row configuration at higher St ($\mbox{St} \geq 0.3$). Increasing wavelength generally enhances thrust production by reducing both pressure and friction \fnl{of} drag components. Additionally, we identify critical vortex interactions between the front and hind flippers, with destructive interference occurring at lower St and constructive patterns emerging at higher \fnl{$\mbox{St}$}. Circulation analysis confirms stronger vortex formation at higher \fnl{$\mbox{St}$ and $\lambda^\ast$}, particularly during the left stroke phase. These results provide novel insights into the hydrodynamic mechanisms underlying seal locomotion and contribute to our understanding of efficient aquatic propulsion systems.

\end{abstract}

\section{Introduction}
\label{sec:Intro}

Aquatic locomotion is a fundamental aspect of \edt{life,} survival\edt{,} and adaptation for a variety of marine organisms, with fish being exemplary models of efficient movement in underwater environments \cite{sfakiotakis1999review}. The ability to maneuver and propel effectively through water not only \edt{aids} these species \edt{to survive in challenging circumstances} by facilitating \edt{avoidance from predators, capturing preys}, foraging, and migration but also serves as an invaluable source of inspiration for \edt{developing new} engineering applications \edt{and technologies}. Fish, having \edt{their physiological structures} evolved over millions of years, exhibit diverse \edt{kinematic} swimming modes that optimize their \edt{interactions} with \edt{the surrounding} fluid\edt{. It makes} them ideal subjects for studying the principles of momentum transfer and force generation in aquatic settings. It \edt{significantly} contributes to the development of bio-inspired underwater robotics \edt{\cite{zhu2019tuna, li2020vortex, liu2024real, qing2024spontaneous}}, where the highly evolved and efficient swimming mechanisms of fish inform the design of agile and versatile robotic systems. For instance, the modulation of \edt{the angular orientation of the fins of the fish \cite{akhtar2007hydrodynamics, liu2017computational, han2020hydrodynamics, khalid2021larger, borazjani2013fish}} and the role of the caudal peduncle \edt{\cite{wang2020tuna, matthews2022role}} in \edt{producing} thrust observed in fish \edt{are} instrumental in improving \edt{propulsive} mechanisms of robotic fish. Secondly, \edt{examining} hydrodynamics of different swimming styles across diverse \edt{marine} species reveals the intricate relationship between \edt{their physiology, undulatory} kinematics, and swimming performance. \edt{These critical elements} elucidates the evolutionary adaptations that enable fish to thrive in \edt{large bodies of water \cite{borazjani2010role, tytell2010disentangling, khalid2021anguilliform, wu2022numerical}}. Beyond biomechanics and fluid dynamics, understanding fish motion enriches our knowledge of fish behavior, ecology, and evolutionary biology. This holistic understanding not only advances biological sciences but also informs conservation strategies and ecosystem management.

\edt{In the last three decades, numerous integrated scientific investigations were conducted and presented by different research teams from the fields of biology and mechanical/marine engineering.} Within the diverse propulsion methods observed in \edt{aquatic animals}, the body-caudal Fin (BCF) and median-paired fin (MPF) modes, as categorized by \edt{Fish} \cite{fish1979form}, represent two major swimming strategies. \edt{The} BCF mode, characterized by undulatory movements along the body and caudal fin, encompasses swimming styles\edt{,} such as thunniform and carangiform, each adapted to different hydrodynamic requirements. \edt{The} carangiform swimming \edt{mode}, for example, involves limited undulation primarily in the posterior half of the body, with a sharp increase in amplitude towards the caudal fin, as detailed by \cite{tytell2010disentangling,cui2017cfd}. This mode strikes a balance between maneuverability and efficiency, making it prevalent among fast-swimming species \cite{borazjani2008numerical,costa2020design}. In contrast, thunniform swimming, utilized by species like tuna and lamnid sharks, restricts undulation to the rear third of the body, achieving high efficiency and speed through a streamlined body and caudal fins \edt{with high aspect ratios} \cite{ben2013exploring,mitin2022bioinspired}. \edt{The major focus of the research efforts remained on carangiform swimmers, mainly driven by complex interactions between their trunks and fins, and anguilliform swimmers \cite{fish2006passive, lauder2007fish, lauder2015fish, lauder2016structure, fish2020bio, amal2024bioinspiration}. Besides, a few studies explained hydrodynamic principles of manta rays, sting rays, and batoid fish \cite{liu2015thrust, menzer2022bio, huang2024hydrodynamic}, which are primarily propelled by their pectoral fins.}

\begin{figure}
\centering
{\includegraphics[width=0.95\textwidth]{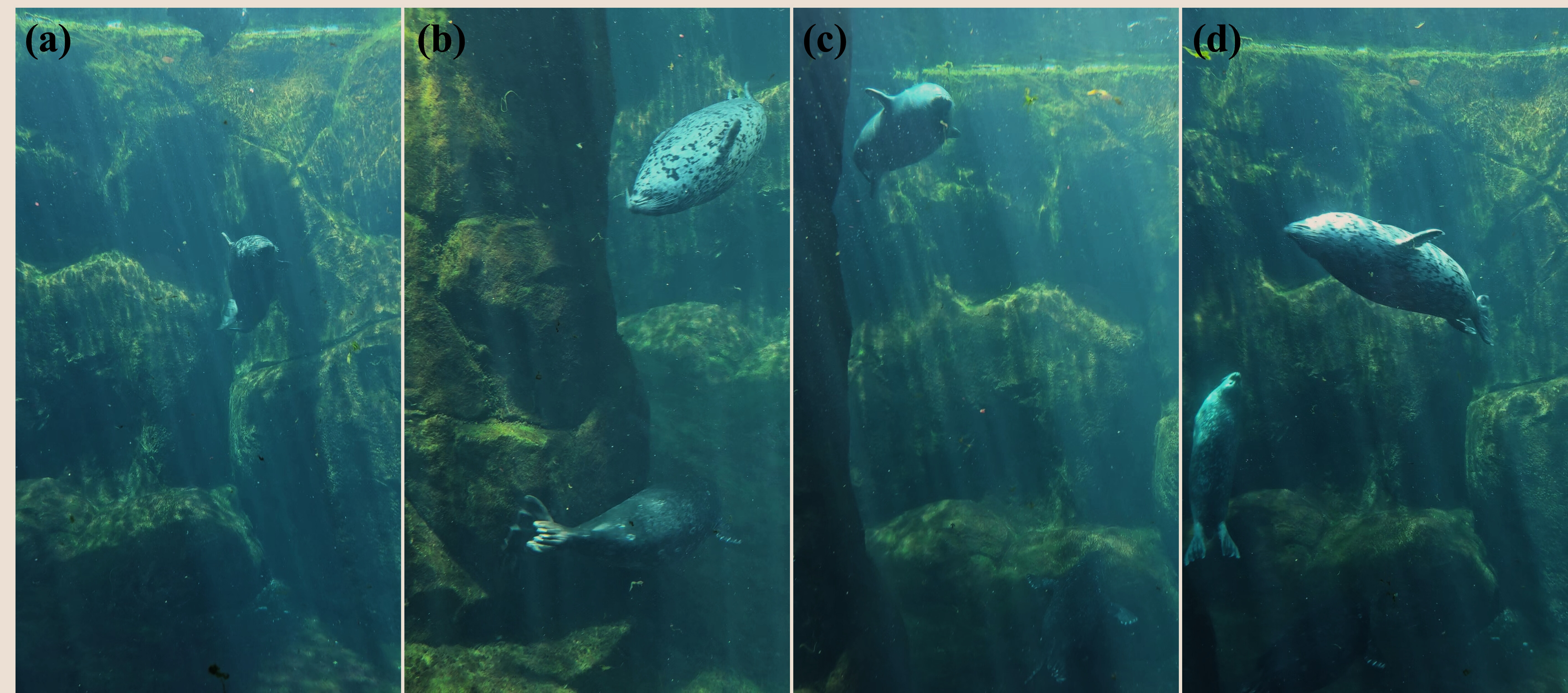}}
\caption{\edt{Images of harbour seals (\textit{Phoca vitulina}) actively using their hind flippers as the main appendages for steady propulsion, captured by MSU Khalid at the Assiniboine Park Zoo, Winnipeg, MB, Canada}}
\label{fig:Seal_Video}
\end{figure}

\edt{Among the diverse population of aquatic marine species, seals or sea lions hold distinct significance due to the shape of their bodies and propulsive mechanisms. There are two distinguished features associated with seals that make them a different kind of swimmer compared to the other well-studies fish: (1) the aspect ratio of their main body, being the main source of frictional drag, and their webbed hind flippers, being one of the primary driving surfaces, is quite different, and (2) they are often observed to steadily swim through their hind flippers that act as dual fins at the posterior part of the body, whereas the other fish mostly have a single caudal fin. Perhaps, Feldkamp \cite{feldkamp1987swimming} and Fish et al. \cite{fish1988kinematics} were among the first investigators to study the kinematics and hydrodynamics of seals, later followed by Stelle et al. \cite{stelle2000hydrodynamic}, Friedman \& Leftwich \cite{friedman2014kinematics}, Perrotta et al. \cite{perrotta2021velocity}, and Leahy et al. \cite{leahy2021role}. Specifically, Feldkamp \cite{feldkamp1987swimming} elaborated the role of fore flippers (pectoral fins) of the California sea lion, \textit{Zalophus californianus}, whereas Fish et al. \cite{fish1988kinematics} mentioned that the hind flippers of harp seals, \textit{Phoca groenlandica Erxleben} and ringed seals, \textit{Phoca hispida Schreber}, propelled \edt{themselves} through their alternate lateral sweeps. Since then, the major focus of research was diverted to the hydrodynamic contributions of fore flippers only, considering them as the sole primary driver of seals in water \cite{friedman2014kinematics, fish2015estimation, kashi2020flowfields, perrotta2021velocity}. Recently, Leahy et al. \cite{liu2024real} examined how hind flippers might play their role to govern the turning maneuvers for California sea lions.

From the work of Fish et al. \cite{fish1988kinematics} and observations made by the authors from the videos of steadily swimming harbour seals (see Fig.~\ref{fig:Seal_Video}), it is evident that their dual hind flippers serve as the active propelling appendages, the hydrodynamics of which, to the best of the authors' knowledge, remained broadly uninvestigated until now. This characteristic feature of seals' hydrodynamics also hold importance, because apparently a set of two small hind flippers are employed to drive comparatively a large body of this marine specie.} In contrast to traditional carangiform fish, the locomotion of harbor seals presents a unique paradigm. Seals, particularly those in the \edt{\textit{Phocidae}} family, utilize their hind flippers that \edt{fundamentally differ} from the \edt{single caudal} fin-based propulsion of fish \cite{fish1988kinematics}. This undulatory swimming mode, while not fitting neatly into the carangiform or thunniform categories, shares more similarities with thunniform propulsors in its efficiency and \edt{mechanisms for production of thrust}. The geometric parameters of fish, such as the aspect ratio ($AR$ = square of the span length divided by the area) of the caudal fin and body fineness ratio, play crucial roles in determining the swimming performance of an animal. An optimal $AR$, typically between that of anguilliform and thunniform swimmers, enhances thrust generation and hydrodynamic efficiency \cite{liu2017computational,zhong2022hydrodynamic}. Fish with low-$AR$ tails typically employ anguilliform swimming with a low \edt{undulating} wavelength, \edt{whereas the} fish with high-$AR$ tails use thunniform swimming \edt{mode} with a \edt{large} wavelength \edt{for their undulation} \cite{borazjani2008numerical}. For \edt{example}, the $AR$ of \edt{a} thunniform swimmer (\textit{Thunnus Obesus}), \edt{a carangiform swimmer (\textit{Trachurus Symmetricus}), and an anguilliform swimmer (\textit{Pomatochistus Minutus}) are $7.48$, $3.6$, and $0.6$, respectively} \ahf{\cite{sambilay1990interrelationships}}. In comparison, the $AR$ of the harbor \edt{seal's hind} flipper is approximately 0.87 \ahf{\cite{sambilay1990interrelationships}}. Interestingly, while the kinematic motion of harbor seals closely resembles that of thunniform and carangiform swimmers, the $AR$ of their their \edt{hind} flippers aligns more closely with that of anguilliform swimmers \cite{sambilay1990interrelationships}. This intriguing contrast highlights the unique \edt{physiological and kinematic characteristics} of seals, blending traits typically associated with different swimming modes. \edt{Additionally}, \edt{the} body fineness ratio (\edt{$FR$} = \edt{the} maximum length to maximum thickness), \edt{residing} between $2$ and $6$, with $4.5$ being theoretically optimal, facilitates streamlined movement and minimal drag (\cite{schakmann2023fish,ohlberger2006swimming}) \edt{for marine species}. Seals possess a streamlined body shape, characterized by a mean fineness ratio of $5.55$ — a value comparable to that of other marine mammals\edt{,} such as dolphins and whales \cite{fish1994influence}. These morphological traits, alongside the complex interplay between body and tail movements, underscore the sophisticated adaptations that enable efficient aquatic locomotion. Moreover, harbor seals operate within a specific range of Strouhal numbers (\edt{$\mbox{St} = {f}{l} / U_\infty$ with $f$ being the oscillation frequency, $l$ the full stroke amplitude of the tail, and $U_\infty$ the free-stream velocity}) that overlap with those of various fish species \cite{fish1988kinematics, eloy2012optimal}, yet their propulsion mechanisms and wake structures are \edt{expected to exhibit} unique characteristics due to \edt{the afore-mentioned important morphological and kinematic features. Moreover, previous studies by Williams et al. \cite{williams1985swimming} and Stelle et al. \cite{stelle2000hydrodynamic} indicated} the hydrodynamic drag and the influence of the body size, shape, and orientation on the swimming performance of pinnipeds, further distinguishing their locomotion from that of fish. These differences in geometry and locomotion mechanics make harbor seals an exceptional model for studying the propulsion and hydrodynamics of aquatic mammals, offering new perspectives and insights that complement our existing knowledge about fish locomotion. \edt{Hence, this discussion sets up the novelty, need, and significance of our present work that focusses on examining the body-flippers interactions in} seals \edt{and the consequent wake structures to determine their hydrodynamic performance}.

\section{Computational Methodology}
\label{sec:Num_Method}
\edt{In this section, we explain the geometry of the swimmer's body, that is a harbour seal, its undulating kinematics, the governing flow equations, the numerical discretization schemes, and the vertification and validation of our in-house computational solver, \textit{VortexDyn}.}

\begin{figure}
    \centering
    \begin{minipage}[c]{0.6\textwidth}  
        \vspace*{\fill}
        \subfloat[]{\includegraphics[width=\textwidth]{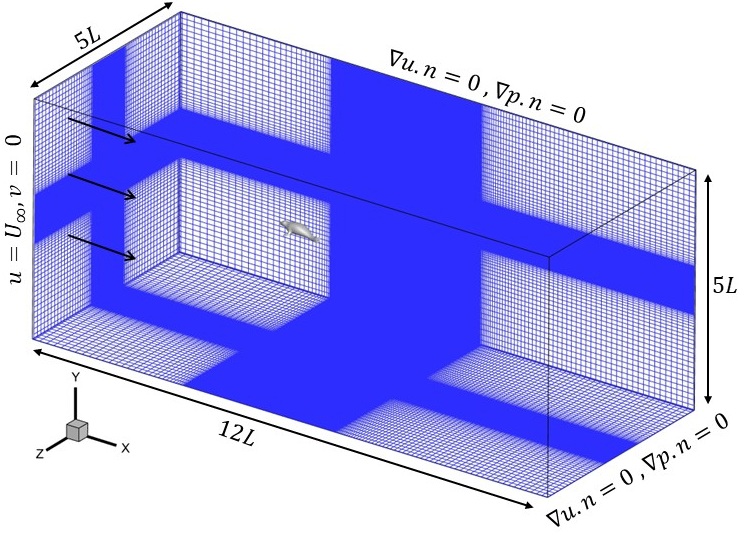}\label{Domain:domain}}  
        \vspace*{\fill}
    \end{minipage}
    \hspace{0.1cm}
    \begin{minipage}[c]{0.35\textwidth}  
        \subfloat[]{\includegraphics[width=\textwidth]{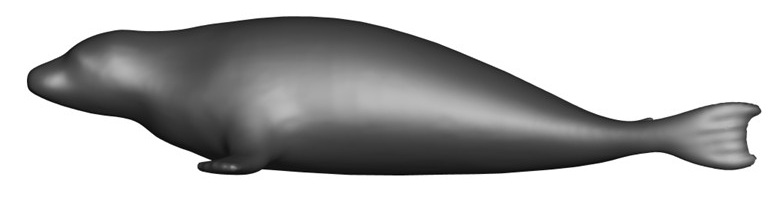}\label{Domain:harbor}}  
        \vspace{0.3cm}  
        \subfloat[]{\includegraphics[width=\textwidth]{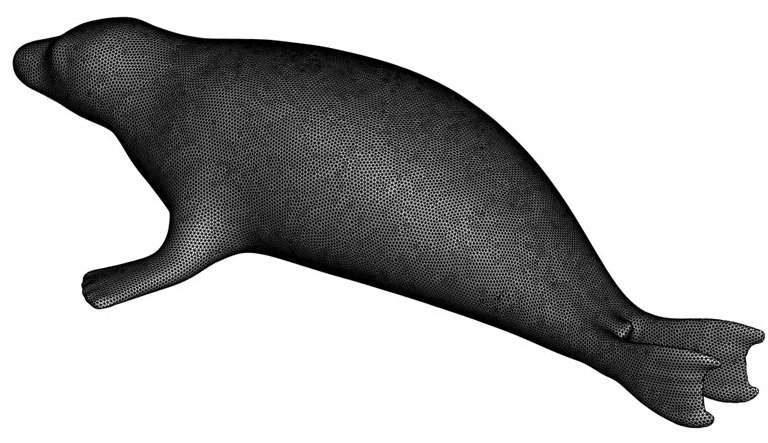}\label{Domain:mesh}}
    \end{minipage}
    \caption{\small\centering (a) Computational domain, (b) side view of harbor seal, (c) surface mesh}
    \label{Domain}
\end{figure}

\edt{
\subsection{Geometry and Kinematics of the Swimmer}
\label{subsec:geom_kin}
For our current study, geometric shape and kinematic profile of a harbour seal is carefully modeled and prescribed} to reflect \edt{its} natural swimming behaviors with its swimming orientation configured based on the descriptions provided by Ul-Haque et al. \cite{ul2015cambered}. From a lateral view, the harbor seal's body \edt{resembles the shape of an airfoil}, that plays a critical role in generating lift and thrust \edt{forces} during swimming. The geometric model, illustrated in Fig.~\ref{Domain:harbor}, highlights the streamlined form and key anatomical \edt{characteristics}. \edt{This physiological model is discretized through triangular mesh cells,} comprising $40,669$ faces and $81,334$ vertex elements to provide a detailed and computationally manageable representation, as presented in Fig.~\ref{Domain:mesh}. 

\edt{In Eq.~\ref{eq:thunniform}, we prescribe the undulatory motion profile of the seal} based on \edt{the} thunniform kinematics\edt{, with the maximum amplitude of $0.26L$ \cite{williams1985swimming,fish1988kinematics}, where $L$ denotes the total length of the \fnl{seal} body}. \edt{Its} amplitude envelope \edt{is} defined using the following relation \ahf{by fitting a \edt{quadratic} polynomial to the local amplitude at given spatial positions along the swimmer's body, nondimensionalized by $L$. The local amplitude used for fitting are A(0)=0.02, A(0.4)=0.01, and A(1.0)=0.13 and is modeled as:}

\begin{align}
A(\frac{x}{L}) &= 0.02 - 0.1150(\frac{x}{L}) + 0.2250(\frac{x}{L})^2; 0 < \frac{x}{L} < 1
\label{eq:thunniform}
\end{align}

\begin{align}
    y(\frac{x}{L}) &= A(\frac{x}{L}) \sin[2\pi({\frac{x}{\lambda^\ast}}-ft)]
\label{eq:motion}
\end{align}

\noindent where $x$ is the stream-wise coordinate, $A$ \edt{represents} the maximum amplitude \edt{of oscillations}, $f$ \edt{shows} the oscillation frequency, \edt{and} $t$ \edt{is} the time. \edt{Besides, $\lambda^\ast$ denotes the non-dimensional} wavelength of the undulatory \edt{motion}. 

\edt{
\subsection{Governing Equations for Fluid Flow and Discretization Schemes:}
\label{subsec:govern_eqn}
}

We conduct three-dimensional ($\mbox{3D}$) numerical simulations \edt{for flows around the undulating} swimmer. The mathematical model for the fluid flow is based on the following non-dimensional forms of the continuity and incompressible Navier-Stokes equations:

\begin{align}
\frac{\partial u_j}{\partial x_j} &=0\\
\frac{\partial u_i}{\partial t} + {u_j}\frac{\partial}{\partial
x_j}({u_i}) &= -\frac{1}{\rho}\frac{\partial p}{\partial
x_i}+ \frac{1}{\mbox{Re}}\frac{{\partial^2}u_i}{{\partial x_j}{\partial x_j}} 
\label{eqn:CartCont}
\end{align}

\noindent where $\{i,j\}=\{1,2,3\}$, $x_i$ shows \edt{the} Cartesian directions, the $u_i$ denotes the Cartesian components of \edt{velocity of} the fluid, $p$ is pressure, and $\mbox{Re}$ represents \edt{the} Reynolds number. \edt{Here, it is defined as $\mbox{Re}={U_\infty}{L}/\nu$, where $\nu$ and $\rho$ show kinematic viscosity of the fluid and its density, respectively}. 


We \edt{solve} these governing equations using a sharp-interface immersed boundary method (IBM) based solver on a non-uniform Cartesian grid, employing radial-basis functions as the interpolation scheme to accurately capture the immersed bodies \cite{farooq4874977accurate}. The IBM serves as \edt{an effective computational} approach\edt{, which simulates complex fluid-structure interactions in challenging scenarios}, such as those \edt{for the harbor seal in the current work}. This method is particularly advantageous in scenarios involving moving boundaries or complex geometrical configurations, where traditional mesh-based approaches may struggle. By utilizing a simplified Cartesian grid, IBM significantly streamlines the mesh generation process, eliminating the need for body-fitted meshes typically required in conventional methods \cite{mittal2008versatile, farooq4874977accurate}. This technique is crucial for handling the complex morphology of aquatic swimmers \cite{khalid2021larger, khalid2021anguilliform, wang2020tuna, zhang2022vortex}. A central difference scheme is used for spatial discretization to approximate the diffusion term, while the convection term is discretized with the Quadratic Upstream Interpolation for Convective Kinematics (QUICK) scheme. The \edt{integration in time} is performed using a fractional-step method, \edt{attaining} second-order accuracy in both time and space. The prescribed wavy motion is imposed as a boundary condition on the swimmer's body, enforced through a ghost-cell technique suitable for both rigid and flexible structures \cite{farooq4874977accurate}. \ahf{Neumann boundary conditions are applied at the far-field boundaries, except at the left-sided inlet boundary, where Dirichlet conditions are specified for the inflow}. Additional details on \edt{our} fully parallelized solver and its application to \edt{various} problems, \edt{relying on complex fluid-structure interactions}, can be found in Ref. \cite{farooq4874977accurate}.

\edt{With the model harbor seal placed in it}, the computational domain \edt{in this work is constructed} as a rectangular volume with dimensions set as $12L$ in \edt{the streamwise direction} and $5L$ \edt{in both vertical and lateral directions}, as exhibited in Fig.~\ref{Domain} which this domain size has also been previously utilized in other studies\cite{khalid2021anguilliform}\cite{liu2017computational}\cite{han2020kinematics}. This sizing \edt{is} carefully chosen to ensure that \edt{the} boundary effects \edt{on \fnl{flow} dynamics around the body} remain negligible, allowing for \edt{accurately capturing the} hydrodynamic interactions during locomotion. A non-uniform structured mesh, comprising approximately $19$ million grid points \edt{is} employed to \edt{optimally using} computational resources while maintaining \edt{a high resolutionin grid in the} regions of interest. The mesh density \edt{is} significantly increased in the immediate vicinity of the \edt{model swimmer} to accurately resolve complex flow features\edt{,} such as boundary layers, shear layers, and \edt{other coherent flow structures}. This refined meshing extends uniformly up to \edt{a distance of} $4L$ downstream of the \fnl{system}. 


To quantify hydrodynamic performance, we \edt{define} the coefficients of drag and lift\edt{, denoted as $C_D$ and $C_L$, respectively} using the following \edt{formulations: $C_D = {F_D} / (0.5{\rho}{{U_\infty}^2}{A_s}$) and $C_L = {F_L} / (0.5{\rho}{{U_\infty}^2}{A_s}$)}, where $F_D$ and $F_L$ represent the drag and lift forces, respectively, and \ahf{$A_s$ represents the surface area of the seal’s body, which is equal to 0.557}. \edt{Lift and drag forces are computed through integrating pressure and shear stress over the whole body of the swimmer. The coefficients, $C_L$ and $C_D$} provide a normalized \edt{metric} of the forces acting on the seal, facilitating comparisons across different flow conditions and geometries.

\subsection{{Validation \& Verification}}
\label{subsec: valid}
\begin{figure}
\centering
{\includegraphics[width=0.8\textwidth]{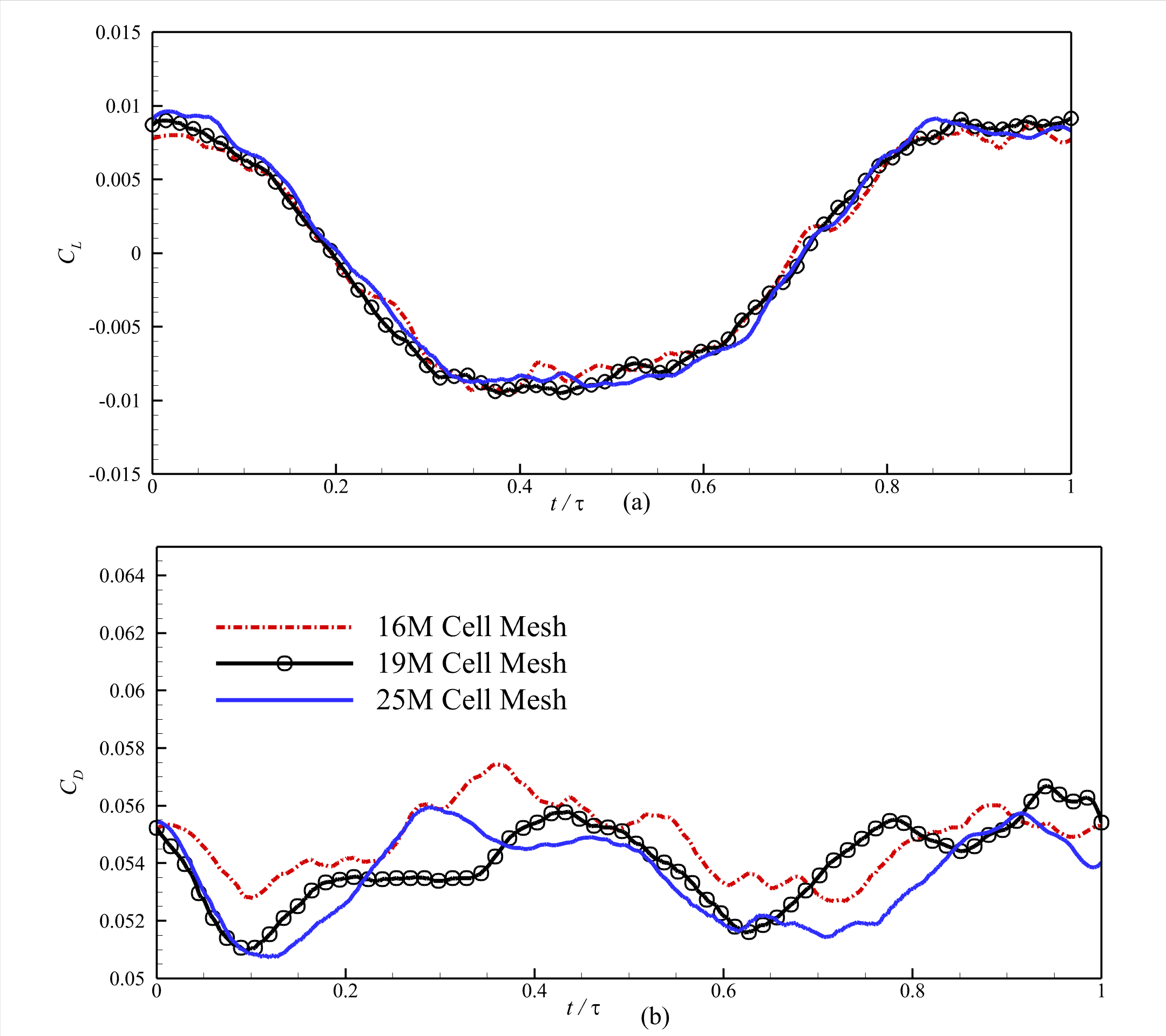}}
\caption{Grid independence analysis for $3$ different grids\edt{, (a) $C_L$ and (b) $C_D$}}
\label{Val_Grid}
\end{figure}

\begin{figure}[htbp]
\centering
{\includegraphics[width=0.8\textwidth]{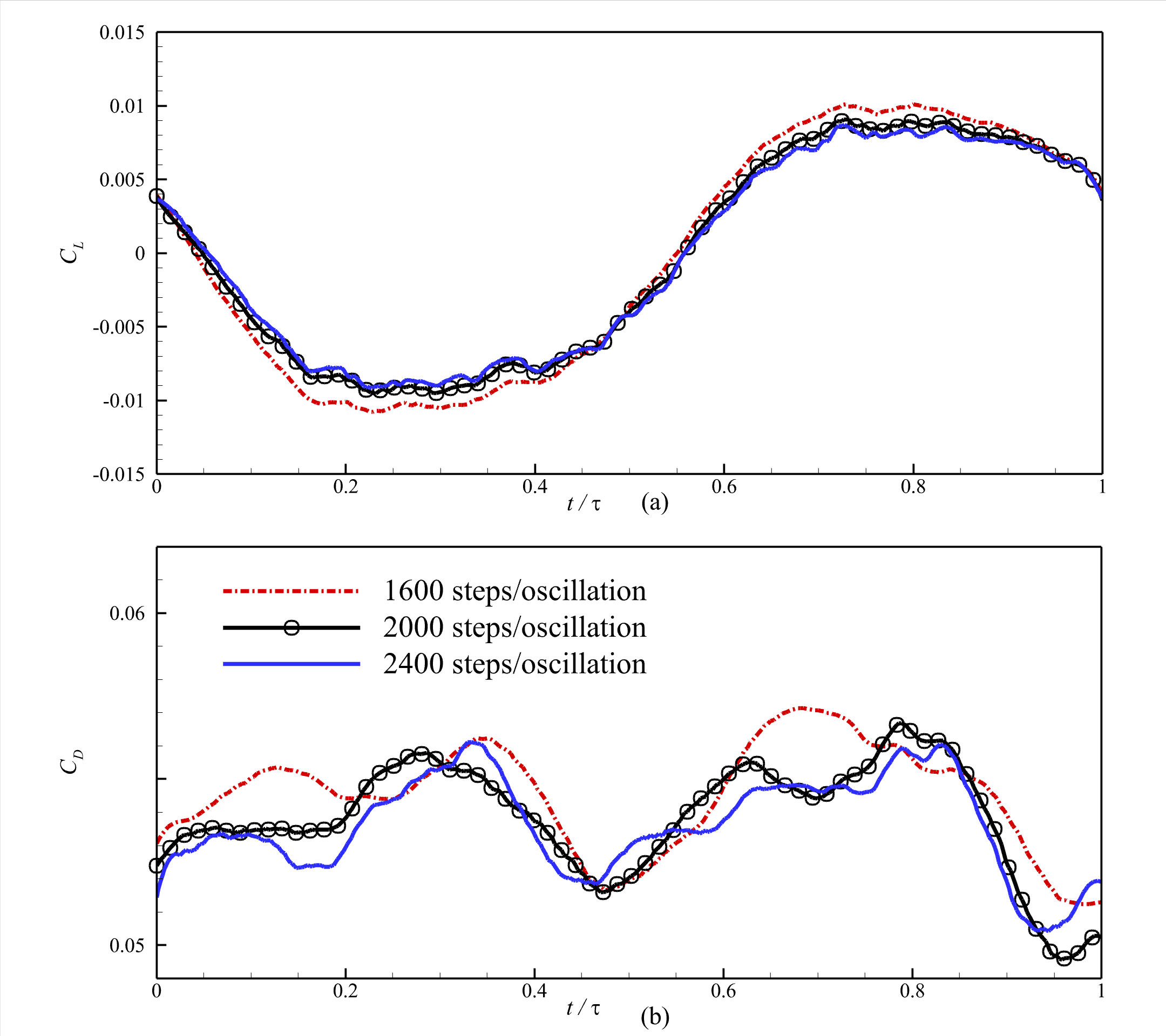}}
\caption{\edt{Results for our time step convergence analysis for $3$ different values of the time steps per oscillation cycle, (a) $C_L$ and (b) $C_D$}}
\label{Val_Time}
\end{figure}

Before conducting the simulations for \edt{our} present study, we \edt{perform} comprehensive grid-independence and time-step convergence studies, as well as validation \edt{tests for our solver for flows around biological model swimmers at $\mbox{Re}=3,000$, \ahf{$\mbox{St} = 0.2$, and $\lambda^\ast = 1.2 $}}. We \edt{carry out} grid-independence tests using three different \edt{mesh} configurations with $2000$ time steps per oscillation cycle. The grid sizes for $\mbox{Grid}~1$, $\mbox{Grid}~2$, and $\mbox{Grid}~3$ \edt{are} approximately $16~\mbox{million}$, $19~\mbox{million}$, and $25~\mbox{million}$ cells, respectively. \edt{Figure}~\ref{Val_Grid} presents comparisons of \edt{temporal profiles of $C_L$ and $C_D$} for one undulation cycle after we attain steady-state solutions in these simulations. Results of the grid-dependence study indicate that increasing the grid resolution from $16~\mbox{million}$ to $25~\mbox{million}$ cells does not result in significant improvements in the computed metrics. While drag (thrust) values do not match perfectly across the grids, the peak values are captured effectively, and further refinement beyond 19 million cells yielded minimal additional improvement. \edt{Table~\ref{tab:sensitivity_analysis} presents error estimates for root-mean square ($\mbox{RMS}$) values of $C_L$ and $C_D$ obtained through the $\mbox{Grid}~1$ and $\mbox{Grid}~2$, considering the finest $\mbox{Grid}~3$ as the reference case. It is evident that the error is minimal for the $\mbox{Grid}~2$ for both hydrodynamic force coefficients}. Consequently, \edt{we choose the $\mbox{Grid}~2$ for numerical simulations in the present study.}. A similar trend \edt{is} observed in the time-step convergence study, \edt{as shown in} Fig.~\ref{Val_Time}\edt. We \edt{employ} $1600$, $2000$, and $2400$ time steps per oscillation cycle. While an improvement \edt{is noticed} when time steps \edt{are increased} from $1600$ to $2000$ \edt{in one undulation cycle, a} further refinement to $2400$ time steps \edt{do} not produce notable \edt{variations}. Therefore, 2000 time steps per oscillation cycle were chosen for the simulations. \edt{Following our approach for the grid-independence study, we determine error estimates for different values of time steps, as shown in Table~\ref{tab:sensitivity_analysis}. These statistical results further justifies our choice of $2000$ time steps per oscillation cycle for our further simulations.}

\begin{table}[ht]
    \centering
    \caption{Grid and Time Step Sensitivity Analysis}
    \setlength{\tabcolsep}{3pt} 
    \begin{tabular}{l|cc|cc||l|cc|cc}
        \toprule
        \multicolumn{5}{c||}{\textbf{Grid Study}} & \multicolumn{5}{c}{\textbf{Time Step Study}} \\
        \midrule
        \makecell{\textbf{Size}\\(Nx, Ny, Nz)} & \textbf{$C_D$} & \textbf{$C_L$} & \textbf{$\Delta C_D$ (\%)} & \textbf{$\Delta C_L$ (\%)} &
        \textbf{Step} & \textbf{$C_D$} & \textbf{$C_L$} & \textbf{$\Delta C_D$ (\%)} & \textbf{$\Delta C_L$ (\%)} \\
        \midrule
        \makecell{16M\\(433, 193, 193)} & 0.0505 & 0.006195 & 11.24 & 3.65 & 1600 & 0.0578 & 0.006541 & 7.45 & 4.8 \\
        \makecell{19M\\(465, 205, 205)} & 0.0550 & 0.006321 & 3.3 & 1.69 & 2000 & 0.0550 & 0.006321 & 2.2 & 1.28 \\
        \makecell{25M\\(513, 225, 225)} & 0.0569 & 0.006430 & -- & -- & 2400 & 0.0538 & 0.006243 & -- & -- \\
        \bottomrule
    \end{tabular}
    \label{tab:sensitivity_analysis}
\end{table}

\begin{figure}
\centering
{\includegraphics[width=0.8\textwidth]{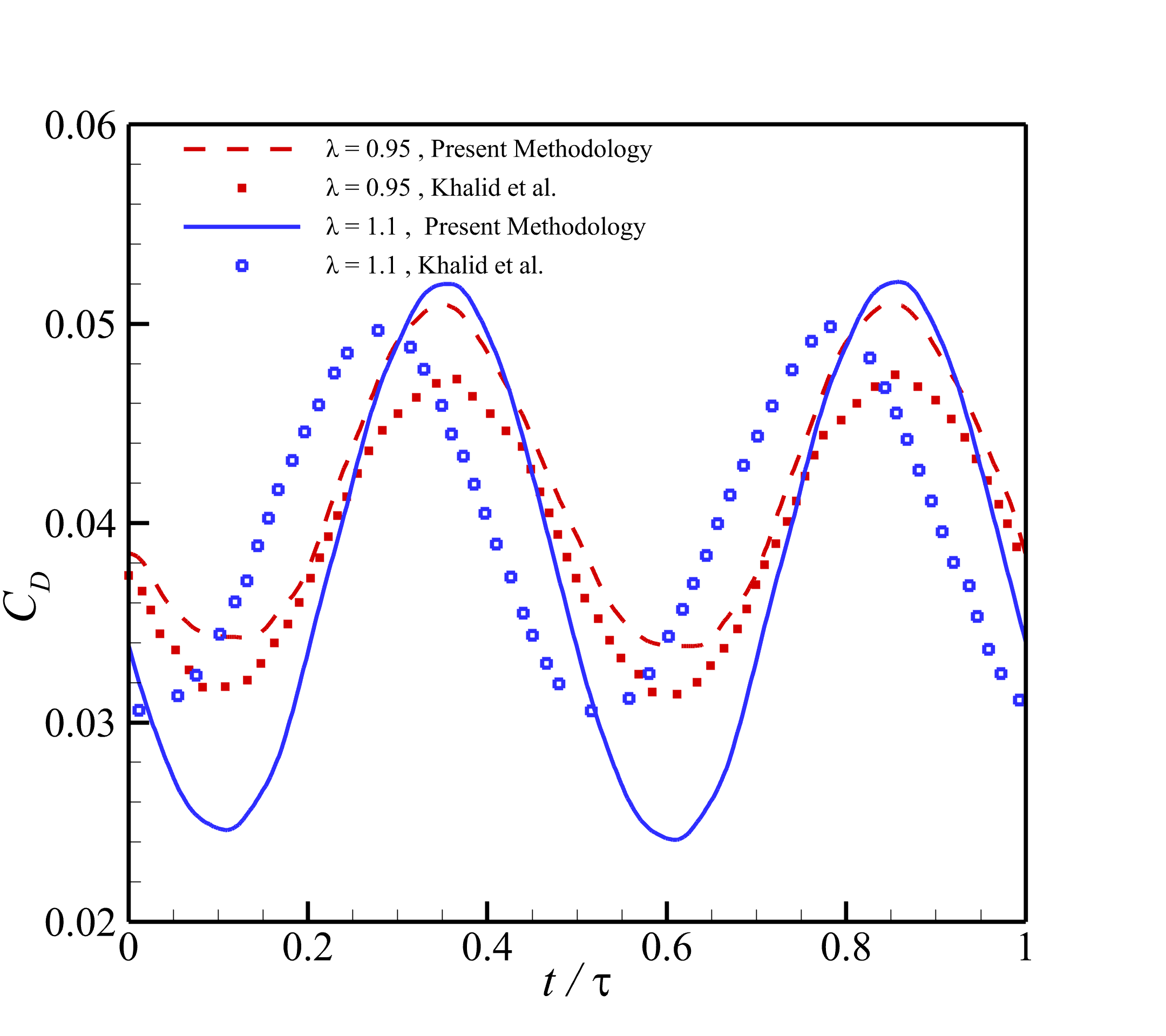}}
\caption{Comparison of the current results with \edt{those} of Khalid et al. \cite{khalid2021anguilliform}}
\label{Eel_Validation}
\end{figure}

\begin{figure}
\centering
{\includegraphics[width=0.8\textwidth]{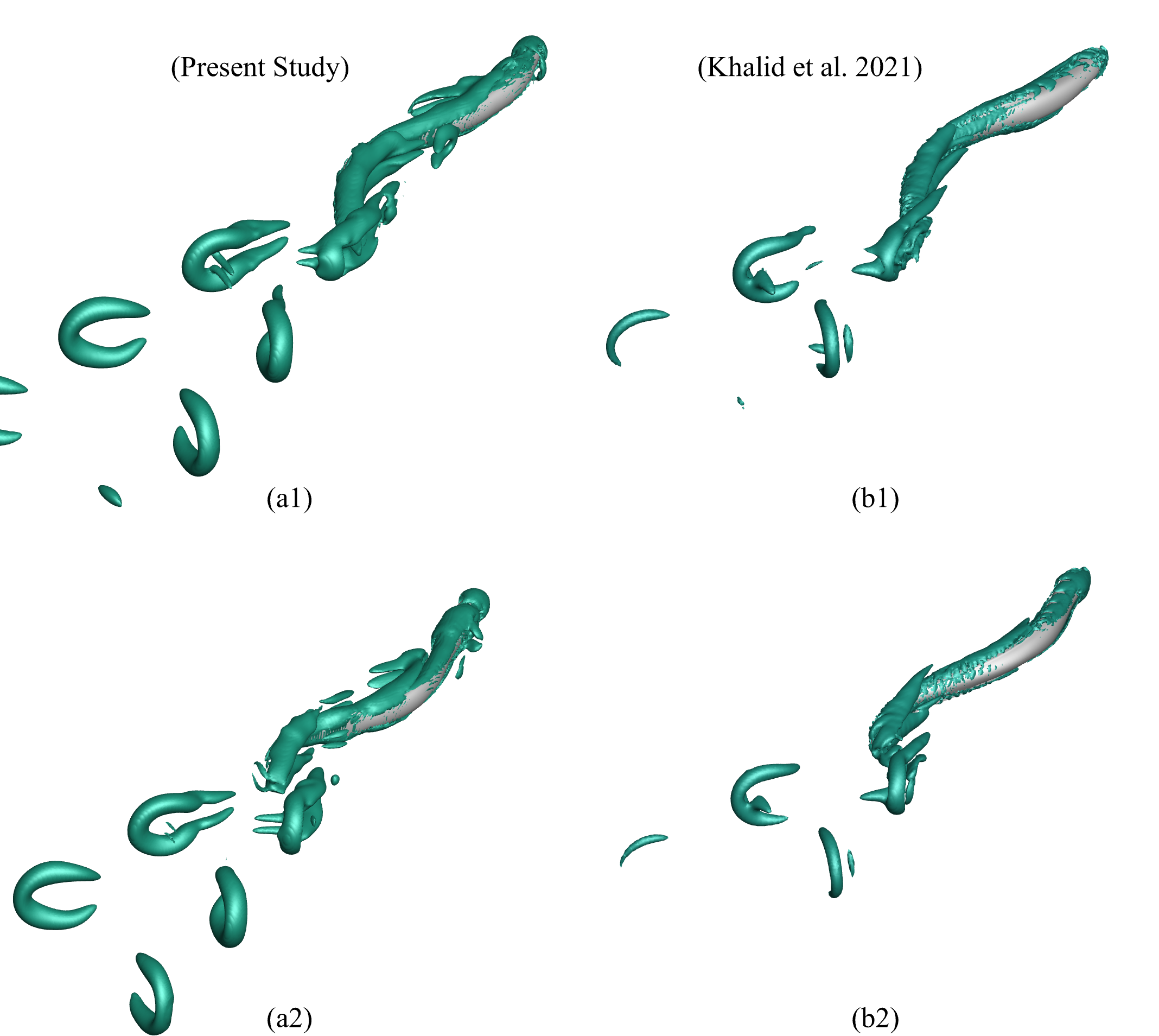}}
\caption{Comparison of the wake structure from the current results with those of Khalid et al. \cite{khalid2021anguilliform} \fnl{for a swimming Eel} \edt{in} the middle (a1,b1) \edt{and at the} end (a2,b2) of the left stroke \edt{for $\mbox{Re}=3,000$, $\mbox{St}=0.4$, and $\lambda^{\ast}=0.95$}}
\label{fig:Eel_Validation_wake}
\end{figure}

\edt{Using the grid and time step chosen through our respective analyses,} we \edt{utilize the work of Khalid et al. \cite{khalid2021anguilliform} as our reference study. We} consider an \edt{eel-like} geometry\edt{,} undergoing steady undulation \edt{using $\lambda^\ast$ of $0.95$ and $1.1$, $\mbox{Re}=3000$, and $\mbox{St}=0.40$. It is important to mention that the geometry of the slender swimmer is slightly different from the one used by Khalid et al. \cite{khalid2021anguilliform} in the context of non-zero thickness of the posterior part of the body. Figure~\ref{Eel_Validation} presents the comparison of $C_D$ obtained through our simulations and those from the reference study. While our solver slightly overpredicts peaks of the force coefficient for the two kinematic conditions, we believe that there could be two potential causes for the small differences observed here: (1) we use a slightly thicker tail to avoid the computations over a membranous body part, (2) we employ radial-basis function-based interpolation technique to identify surface of the body, that is a nonlinear and more robust technique compared to the bi-linear/tri-linear interpolation method used \fnl{by} Khalid et al. \cite{khalid2021anguilliform}. Nevertheless, in order to demonstrate the effectiveness of our solver, we also present a comparison of the vortex dynamics around the two bodies in \edt{Fig.~\ref{fig:Eel_Validation_wake} at two different time instants during the left stroke of their posterior parts. We observe that coherent flow structures are very similar in their shapes. However, the notable difference is in the resolution of flow features in the wake and around the anterior part of the body. We notice the presence of two structures around mid section of the eel's body in our simulation, whereas they are absent in the work of Khalid et al. \cite{khalid2021anguilliform}. Besides the current simulation methodology provides more resolved wake in farther regions, which is expected to influence pressure and, hence, the computations of force coefficients around the body.} Further validation tests, relating to complex fluid-structure interactions in different scenarios, including biological swimmers, galloping, and flutter, from our solver can be found in Ref. \cite{farooq4874977accurate}. Additionally, more results for other multi-physics simulations for flows around undulating bodies are also reported by Kamran et al. \cite{kamran2024does}.
}

\section{Results and Discussion}
\label{sec:results}


\edt{In this work, we perform our numerical simulations at $\mbox{Re}=3000$, whereas \mbox{St} ranges from $0.20-0.35$ and $\lambda^\ast$ takes the values of $1.0$, $1.1$, and $1.2$, following the observations of Fish et al. \cite{fish1988kinematics}. } Although real harbor seals operate at \ahf{higher Reynolds numbers ($10^4$ to $10^5$)} \ahf{\cite{fish1988kinematics}}, our focus is on capturing the fundamental flow phenomena observed during \edt{their} locomotion while keeping the St within the same range as that of the seal locomotion. Previous studies, such as \edt{from Zhong et al.} \cite{zhong2019dorsal}, demonstrated that \edt{primary vortex} dynamics \edt{persisted} even when there \edt{was} an order-of-magnitude difference \edt{in $\mbox{Re}$ during experiments and computational simulations}. Specifically, Zhong et al \cite{zhong2019dorsal} compared \edt{their} experimental results at $\mbox{Re}$ \fnl{of} $41,000$ to $98,400$ \edt{using an IBM-based} simulations at $\mbox{Re} = 2100$. \edt{They found} that essential aspects of dorsal-caudal fin interactions, including the dorsal fin-induced crossflow and its stabilizing effect on the leading-edge vortex, remained consistent. Based on these findings \edt{and those made by Liu et al. \cite{liu2017computational}}, we \edt{select} $\mbox{Re} = 3000$ for our \edt{current} numerical simulations.



\subsection{Hydrodynamic Performance Metrics}
\label{subsec:hydro}

\begin{figure}[h!]
    \centering
    \begin{minipage}[b]{0.32\linewidth}
        \centering
        \subfloat[]{\includegraphics[width=\linewidth]{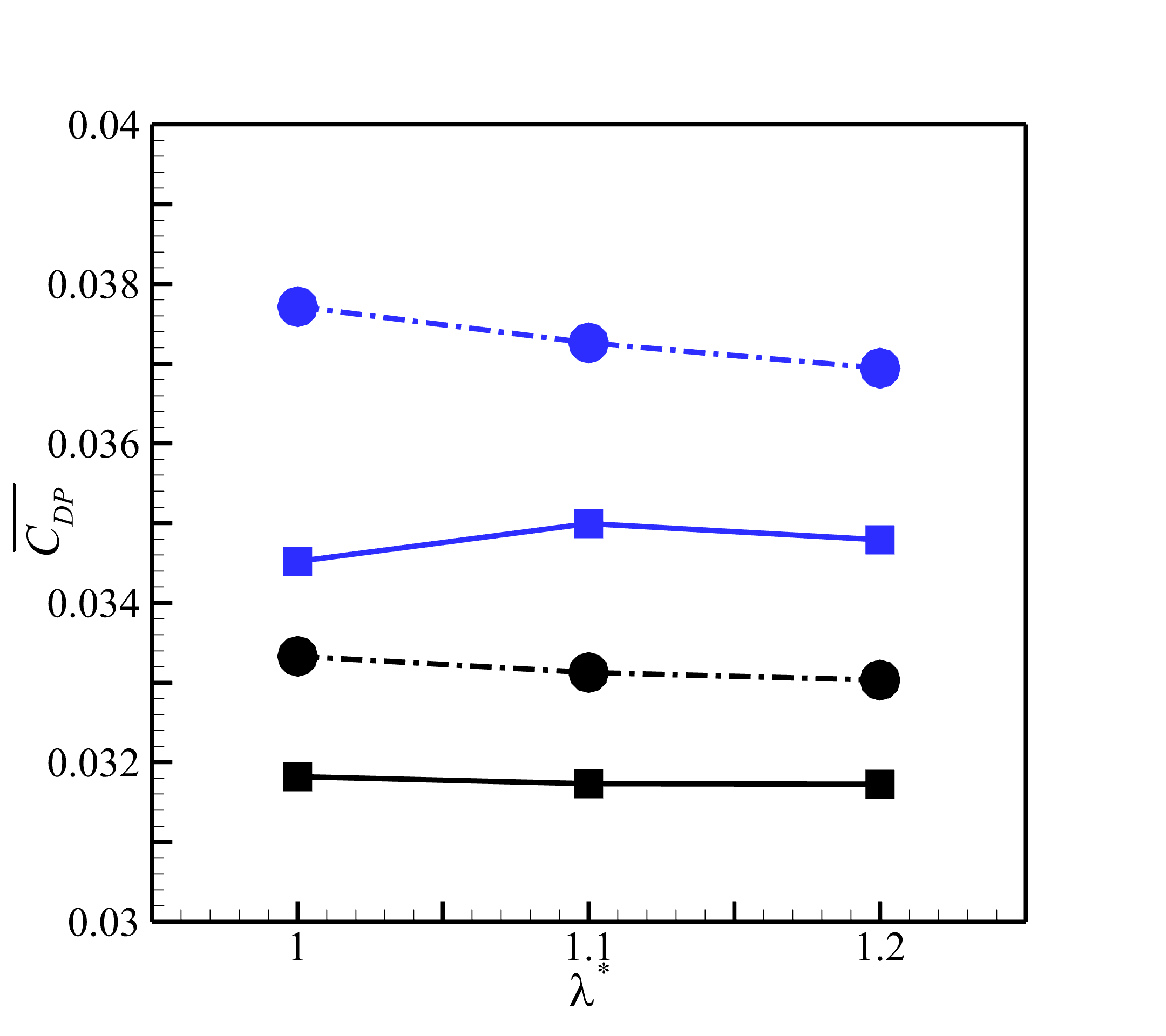}\label{Ave:CDP}}
    \end{minipage}
    \begin{minipage}[b]{0.32\linewidth}
        \centering
        \subfloat[]{\includegraphics[width=\linewidth]{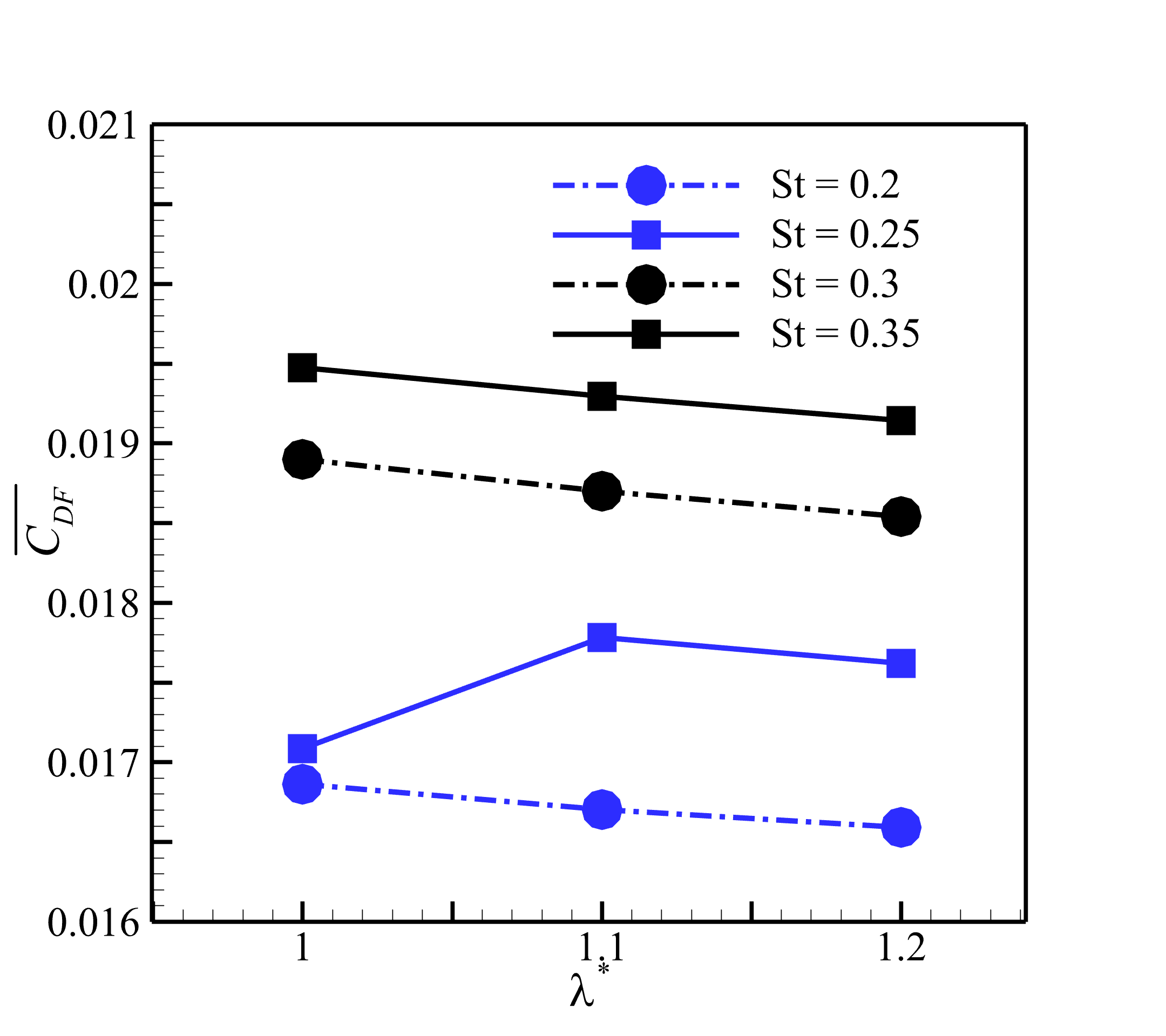}\label{Ave:CDF}}
    \end{minipage}
    \begin{minipage}[b]{0.32\linewidth}
        \centering
        \subfloat[]{\includegraphics[width=\linewidth]{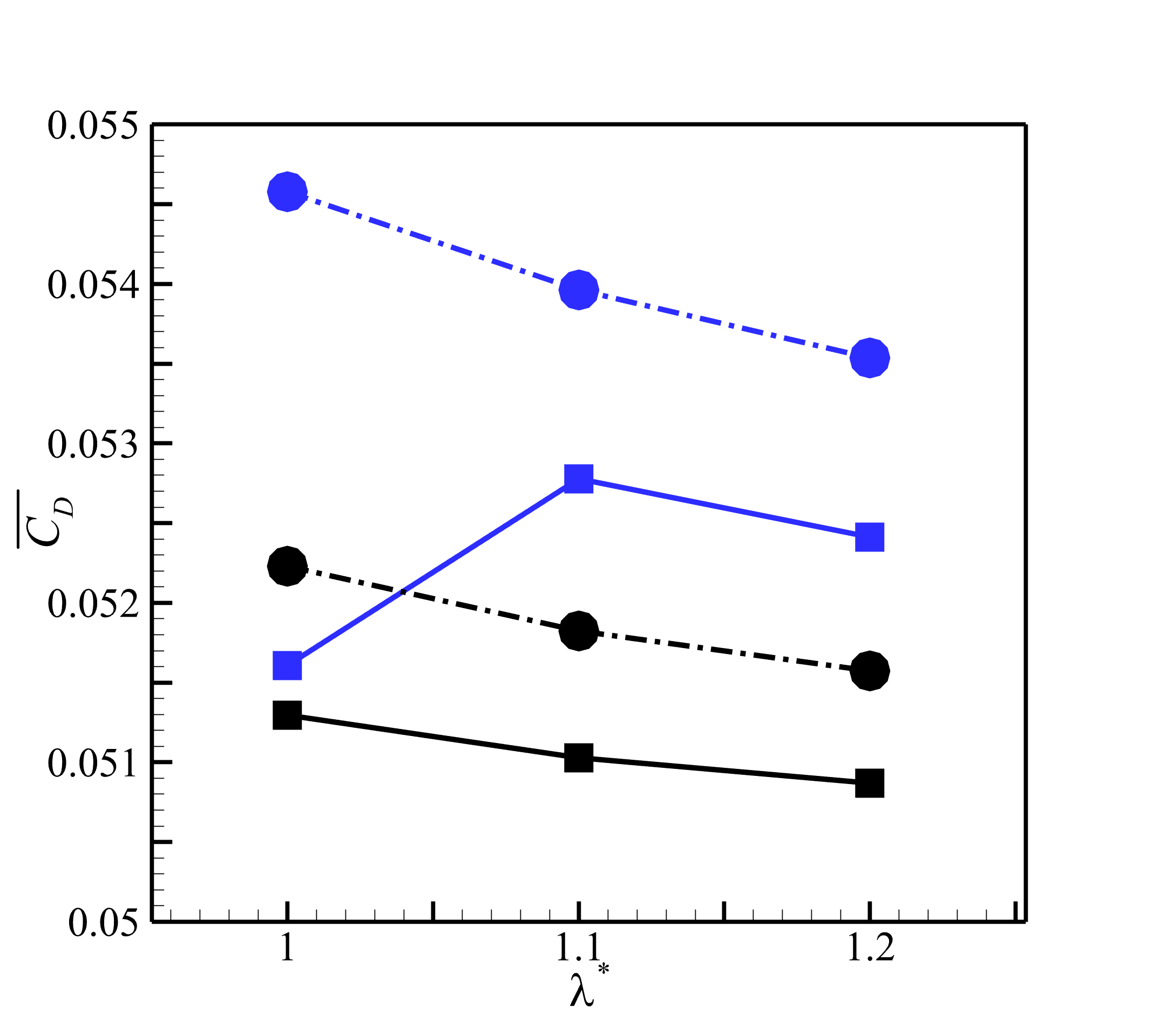}\label{CD}}
    \end{minipage}
    \caption{\small\centering Average drag coefficient and its components for all cases}
    \label{Average_CD_Component_All_Cases}
\end{figure}

\edt{Figure}~\ref{Average_CD_Component_All_Cases} illustrates \edt{how $\overline{C_D}$ and its frictional and pressure components \fnl{of the undulating swimmer}, represented by $\overline{C_{DF}}$ and $\overline{C_{DP}}$, respectively, vary with respect to $\mbox{St}$ and $\lambda^\ast$}. \ahf{Total $\overline{C_D}$ exhibits a consistent general trend across all cases with $\overline{C_D}$ predominantly decreasing \fnl{with} both St and $\lambda^\ast$ increase. This overall reduction in drag suggests an enhancement in swimming \edt{performance} at higher \edt{undulating wavelengths}, particularly when coupled with higher \edt{values of} $\mbox{St}$.} Specifically, at $\mbox{St}=0.25$, an initial increase in $\overline{C_D}$ is observed as $\lambda^\ast$ increases from $1.0$ to $1.1$. However, further increasing $\lambda^\ast$ to $1.2$ leads to a subsequent \edt{reduction} in $\overline{C_D}$. \edt{Figures}~\ref{Ave:CDP} \edt{and} ~\ref{Ave:CDF} \edt{provides insights into the variations of $\overline{C_{DF}}$ and $\overline{C_{DP}}$ at different $\mbox{St}$ and $\lambda^\ast$}. \ahf{When focusing on the effect of $\mbox{St}$, raising it lowers the overall $\overline{C_D}$, even though $\overline{C_{DF}}$ increases. The main driver behind this net improvement is the simultaneous drop in $\overline{C_{DP}}$, hinting that the hind flipper's faster beat frequencies strengthen the seal’s pressure‐based thrust enough to overcome added friction. \edt{It} also suggests that the wake structure becomes more favorable for thrust generation as $\mbox{St}$ increases. More \edt{specifically}, one can observe that at a fixed $\mbox{St}$, increasing $\lambda^{\ast}$ leads to a reduction in \edt{frictional drag}.} \ahf{\edt{This phenomenon is contrary to what is observed for} anguilliform swimmers like eels \cite{khalid2021anguilliform}, which prefer to swim with wavelengths shorter than their body lengths. Therefore, our findings indicate that harbor seals favor higher wavelengths for swimming.}

\begin{figure}[h!]
    \captionsetup[subfigure]{labelformat=empty} 
    \centering
    \begin{minipage}[b]{0.4\linewidth}
        \centering
        \subfloat[(a1)]{\includegraphics[width=\linewidth]{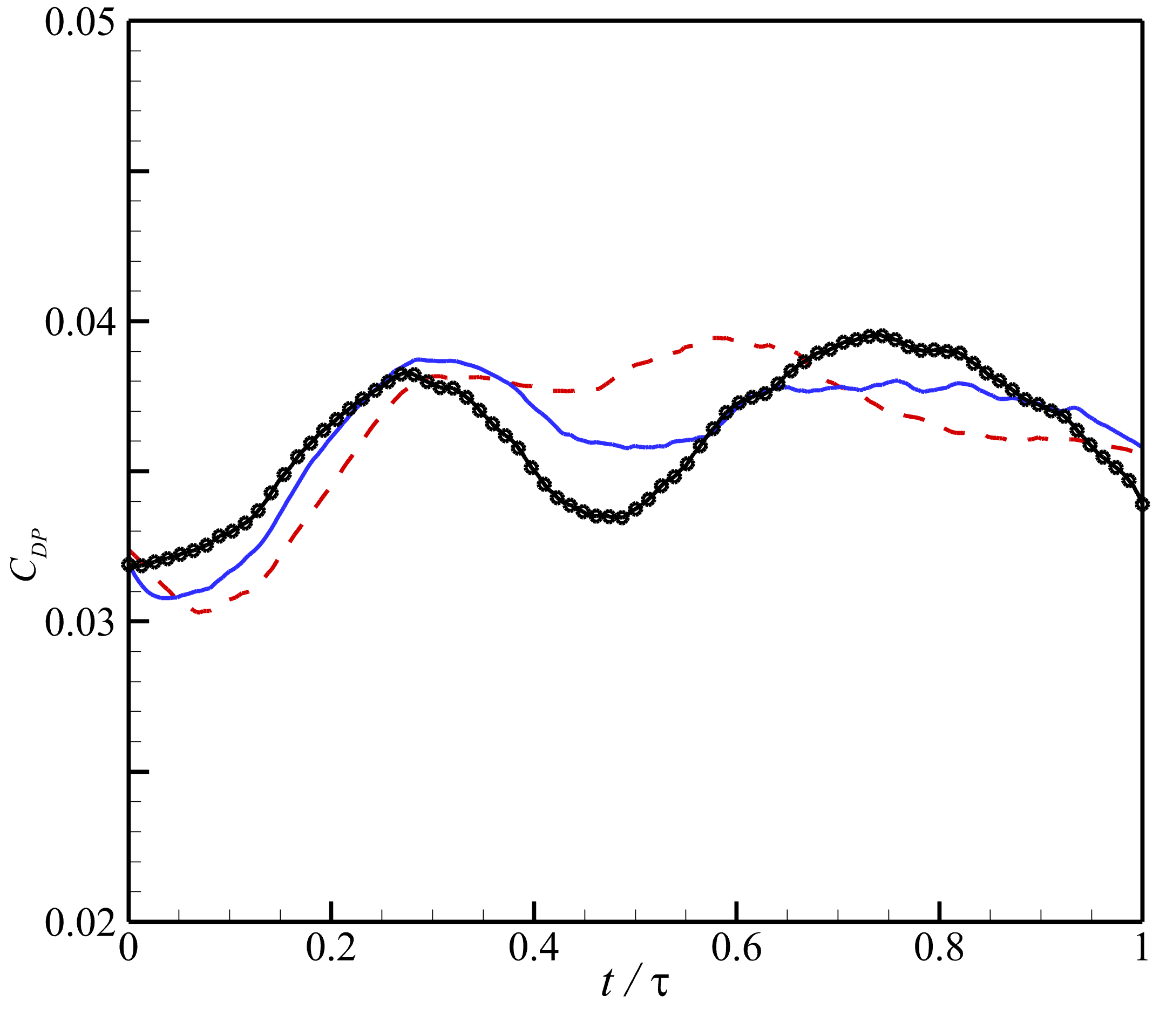}}
    \end{minipage}
    \begin{minipage}[b]{0.4\linewidth}
        \centering
        \subfloat[(b1)]{\includegraphics[width=\linewidth]{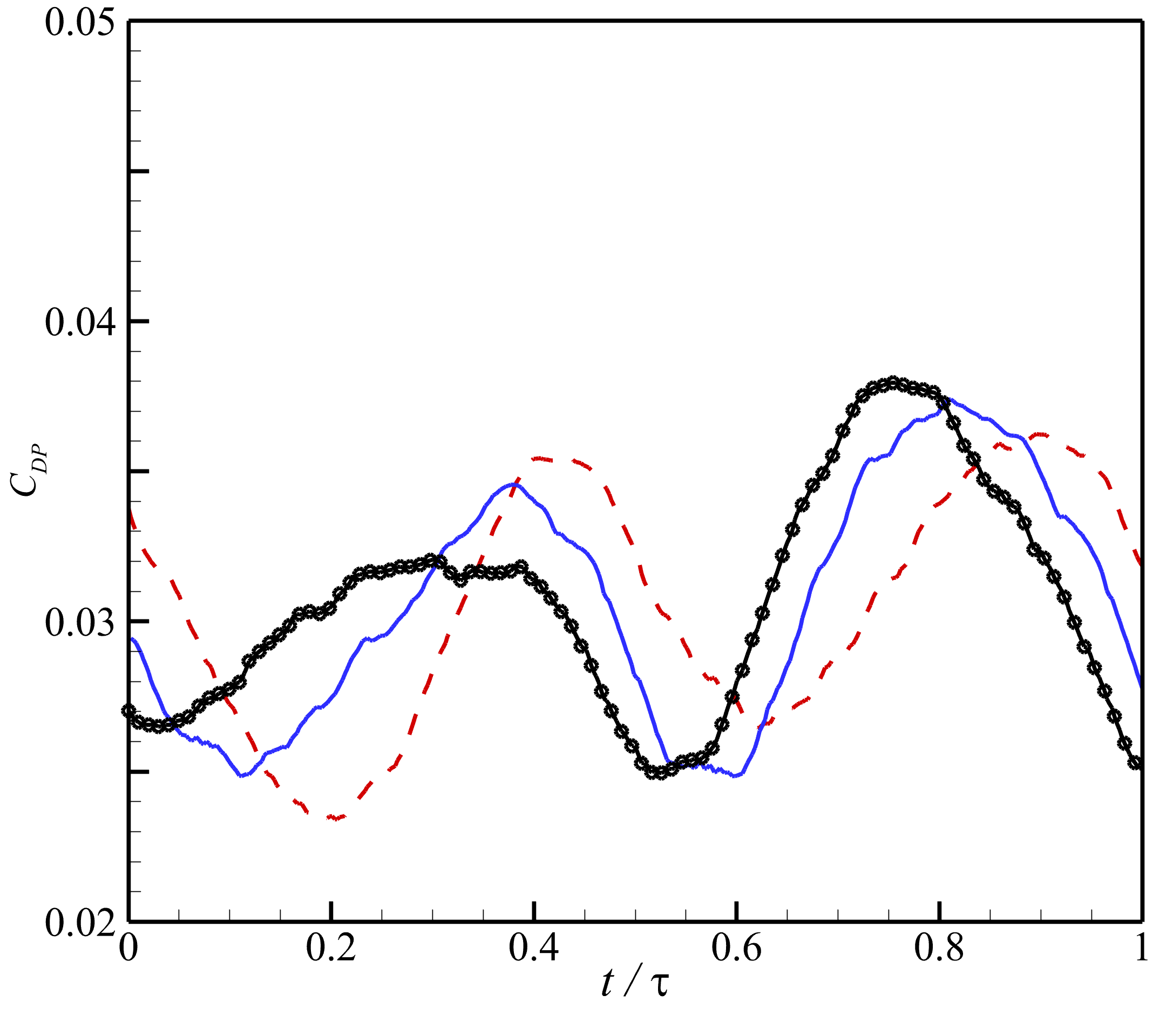}}
    \end{minipage}
    
    \vspace{0.0cm}
    
    \begin{minipage}[b]{0.4\linewidth}
        \centering
        \subfloat[(a2)]{\includegraphics[width=\linewidth]{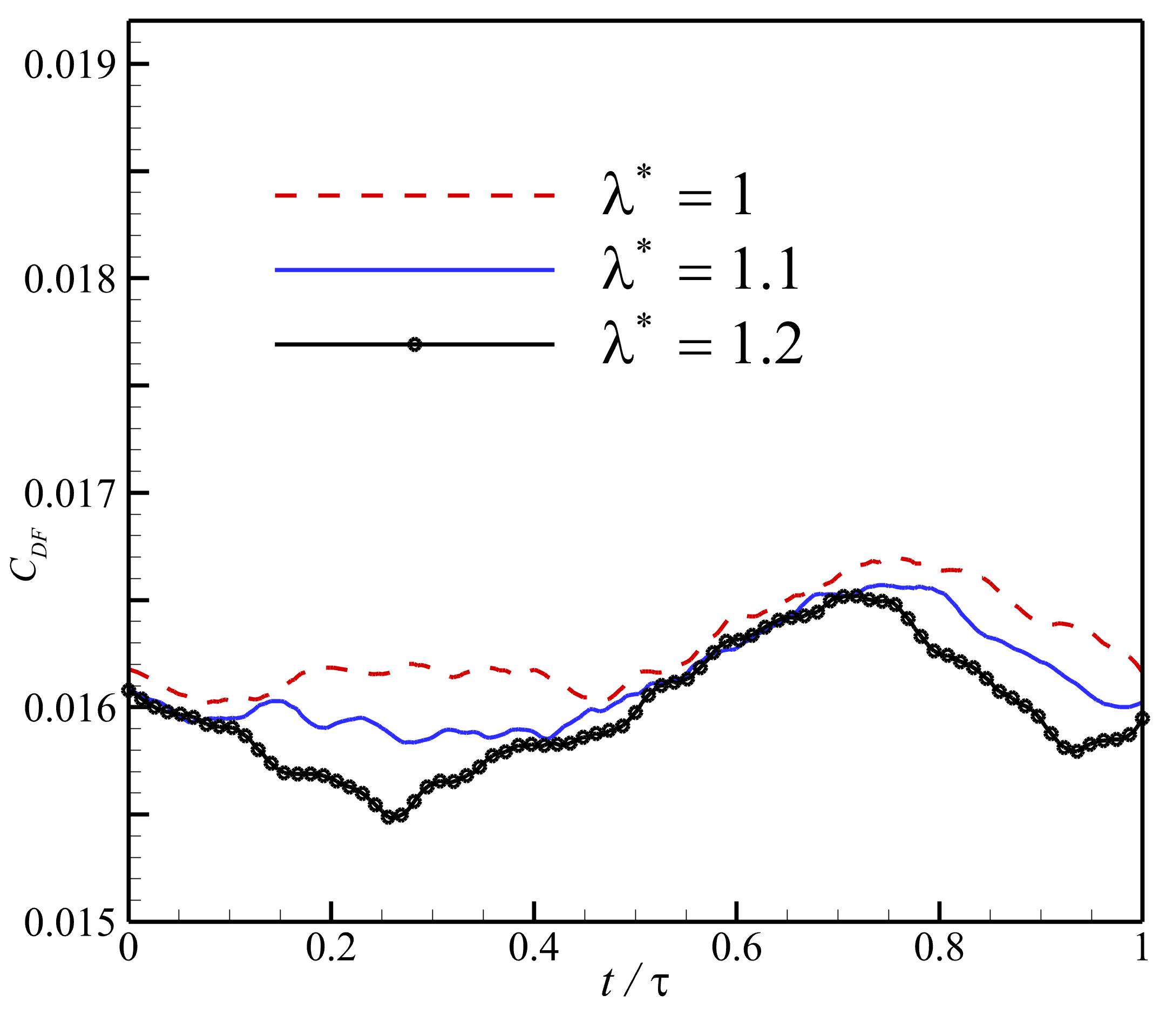}}
    \end{minipage}
    \begin{minipage}[b]{0.4\linewidth}
        \centering
        \subfloat[(b2)]{\includegraphics[width=\linewidth]{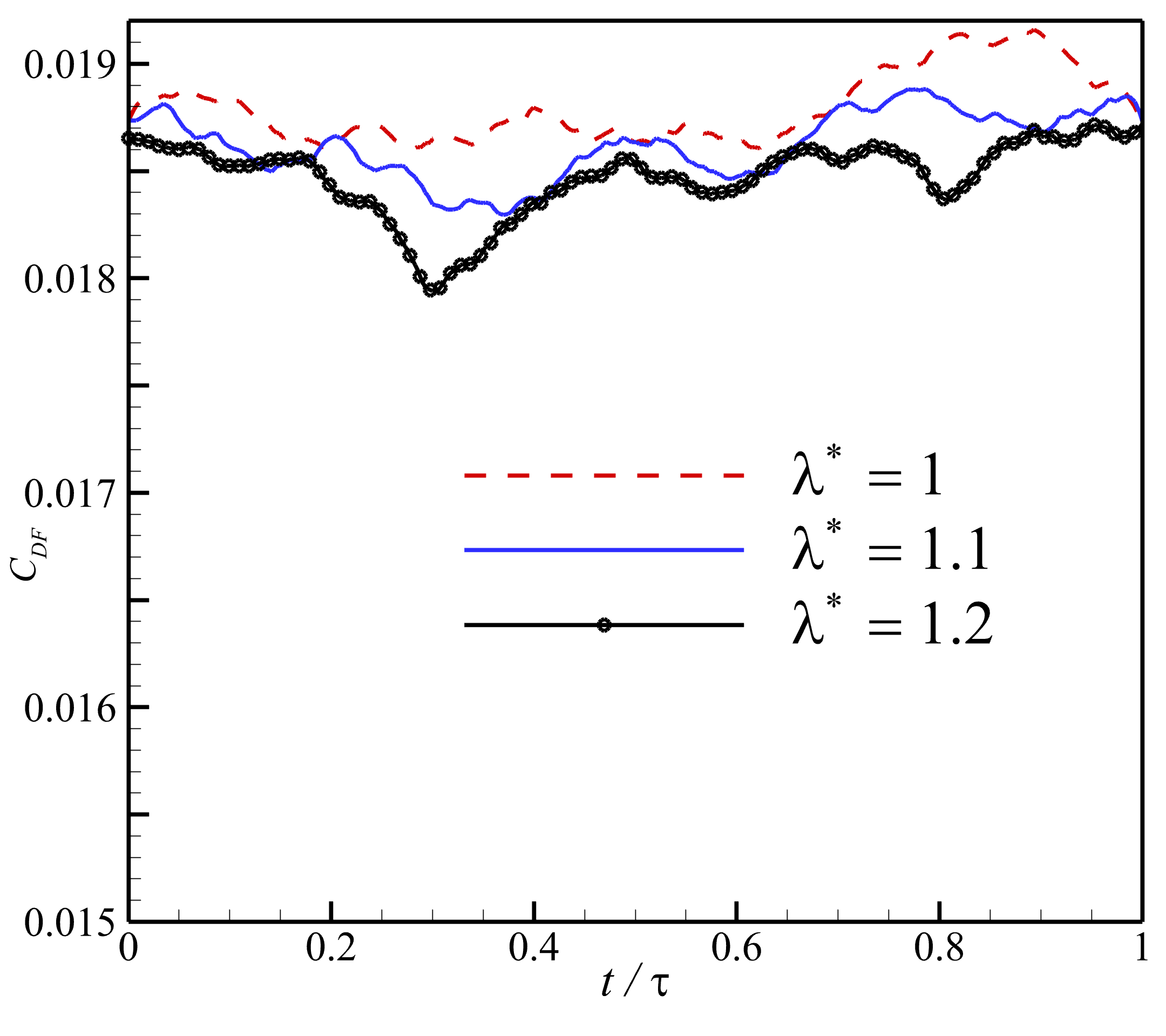}}
    \end{minipage}
    
    \vspace{0.cm}
    
    \begin{minipage}[b]{0.4\linewidth}
        \centering
        \subfloat[(a3)]{\includegraphics[width=\linewidth]{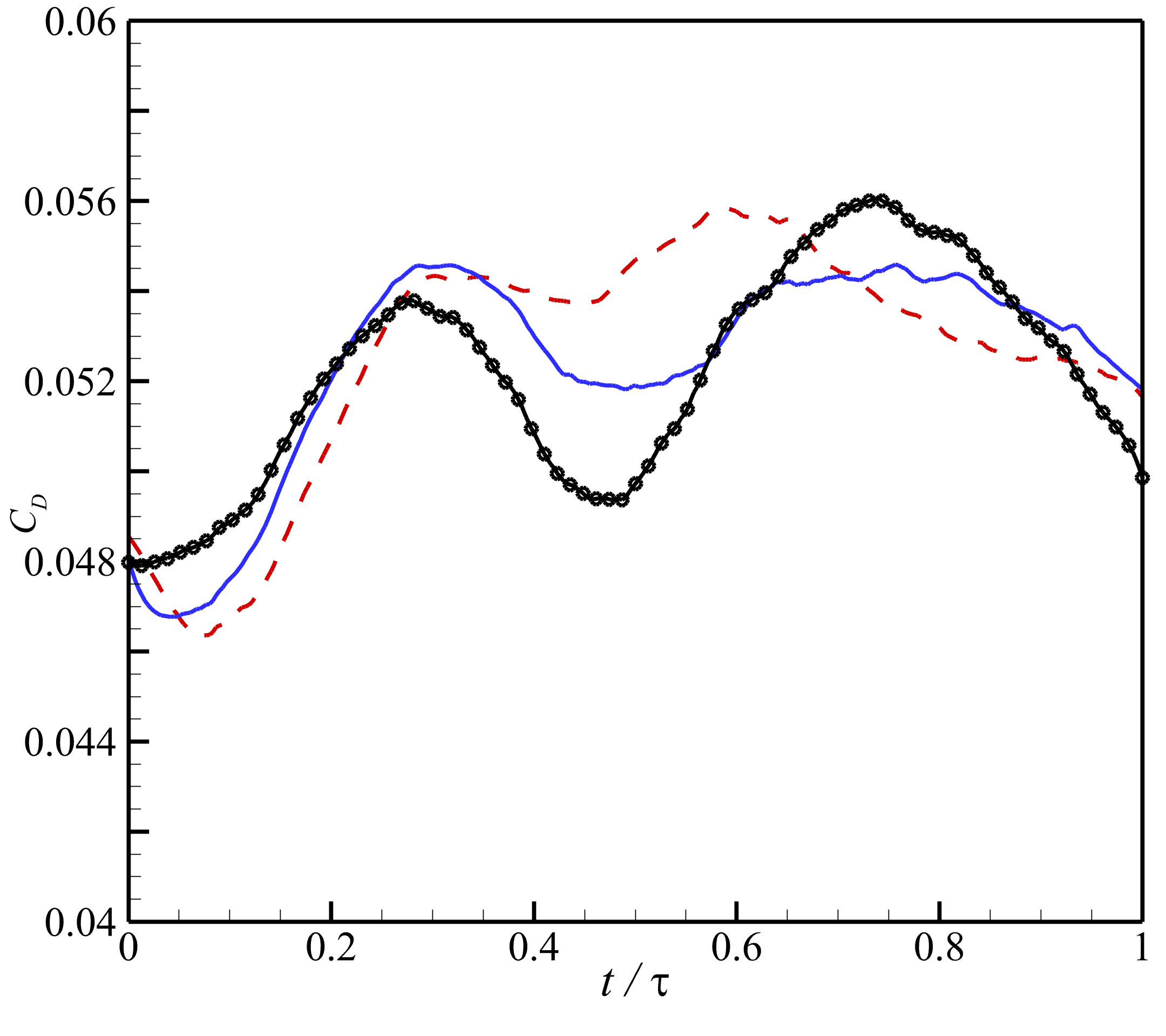}}
    \end{minipage}
    \begin{minipage}[b]{0.4\linewidth}
        \centering
        \subfloat[(b3)]{\includegraphics[width=\linewidth]{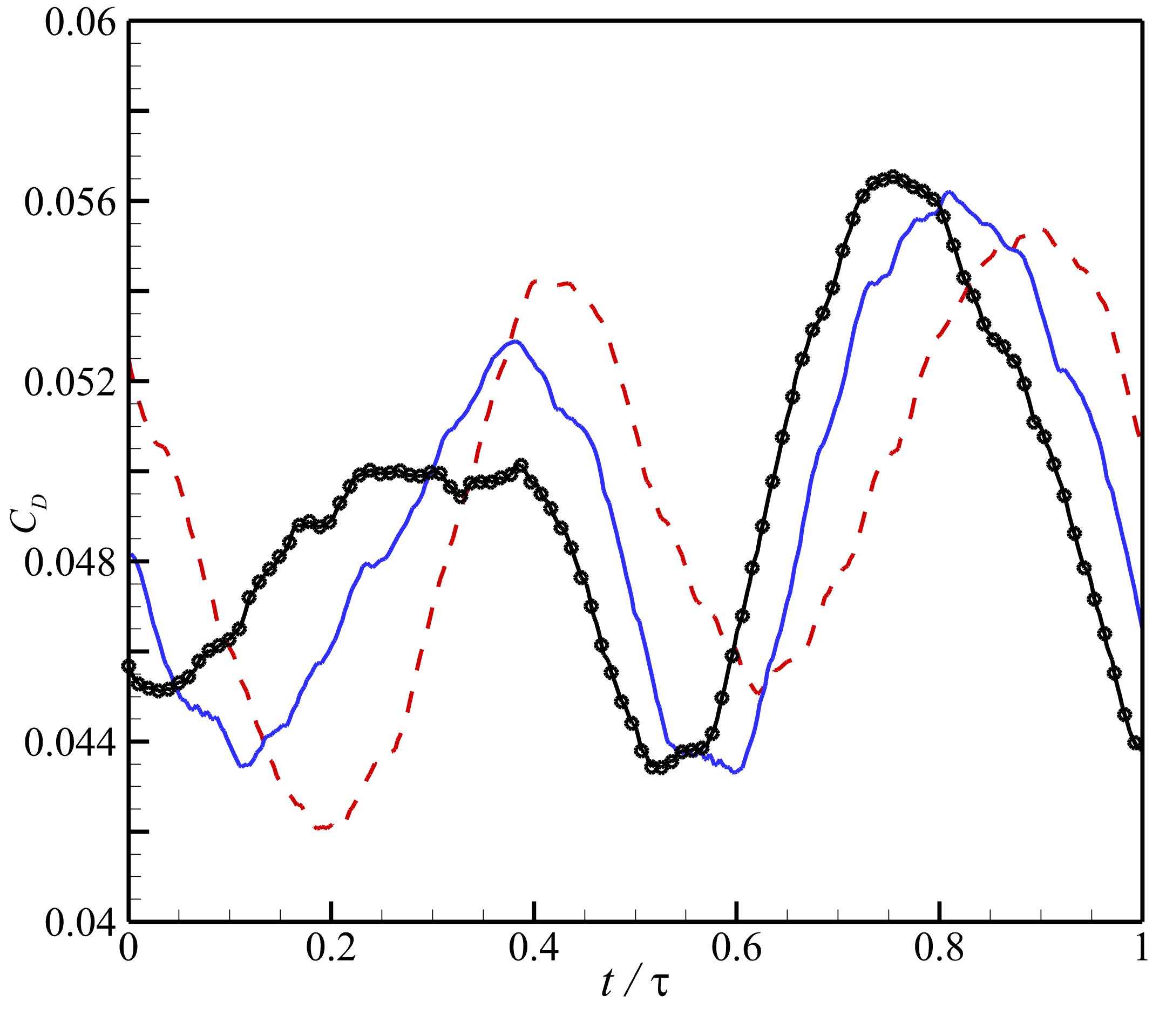}}
    \end{minipage}
    \caption{\small\centering Temporal variations in $C_D$ and its component. (a) for St = 0.2 and (b) for St = 0.35}
    \label{Temporal_CD_component}
\end{figure}

\ahf{Figure~\ref{Temporal_CD_component} presents a detailed view of the temporal variations of $C_D$, $C_{DP}$, and $C_{DF}$ over a single undulation cycle at $\mbox{St}=0.2$ and $0.35$, for varying $\lambda^\ast$. Unlike Fig~\ref{Average_CD_Component_All_Cases} , which presents averaged values, Fig~\ref{Temporal_CD_component} allows us to observe instantaneous fluctuations. A consistent observation \edt{for the two} St \edt{values} is the reduction in the magnitude of $C_{DF}$ \fnl{fluctuations with increasing $\lambda^\ast$}. It indicates that longer wavelengths not only reduce the average frictional drag but also lead to a smoother frictional force profile over the undulation cycle. \edt{On the contrary,} $C_{DP}$ exhibits more pronounced and sharper fluctuations compared to $C_{DF}$. These fluctuations in $C_{DP}$ are directly related to the pressure distribution around the seal's body as it undulates, and are \edt{ultimately} linked to the formation and shedding of vortices. \edt{Besides,} $C_{DP}$ also \edt{exhibits} a trend towards lower magnitudes and potentially a shift in phase with increasing $\lambda^\ast$, contributing to the overall decrement of the time-averaged $C_D$.}

\edt{
\subsection{Formation and Dynamics of Coherent Flow Structures From the Body-Flippers Interactions}
\label{subsec:wake}
}

\edt{In this subsection, we present our analysis for vortex dynamics originated from the body-flippers interactions to determine the hydrodynamics performance of a marine swimmer.} For this purpose, \edt{we employ the} $Q$-criterion to \edt{identify and visualize} coherent \edt{flow} structures \edt{around the swimmer and its wake}. It is defined as $Q = 0.5(\|\omega \|^2  - \|S\|^2)$\edt{,} where $\omega$ is the antisymmetric part (vorticity tensor), and $S$ is the symmetric part (strain rate tensor) of the velocity gradient tensor. Positive values of $Q$ signify regions where the rotation rate dominates the strain rate, indicating the presence of vortical structures. \ahf{Building upon this analysis, we aim to address two key \fnl{aspects}: \edt{(1)} to understand the reasons behind the emergence of different wake patterns, and (2) to evaluate the impact of each distinct pattern on thrust production.}

\begin{figure}[h!]
    \captionsetup[subfigure]{labelformat=empty} 
    \centering
    \begin{minipage}[b]{0.4\linewidth}
        \centering
        \subfloat[(a1)]{\includegraphics[width=0.7\linewidth]{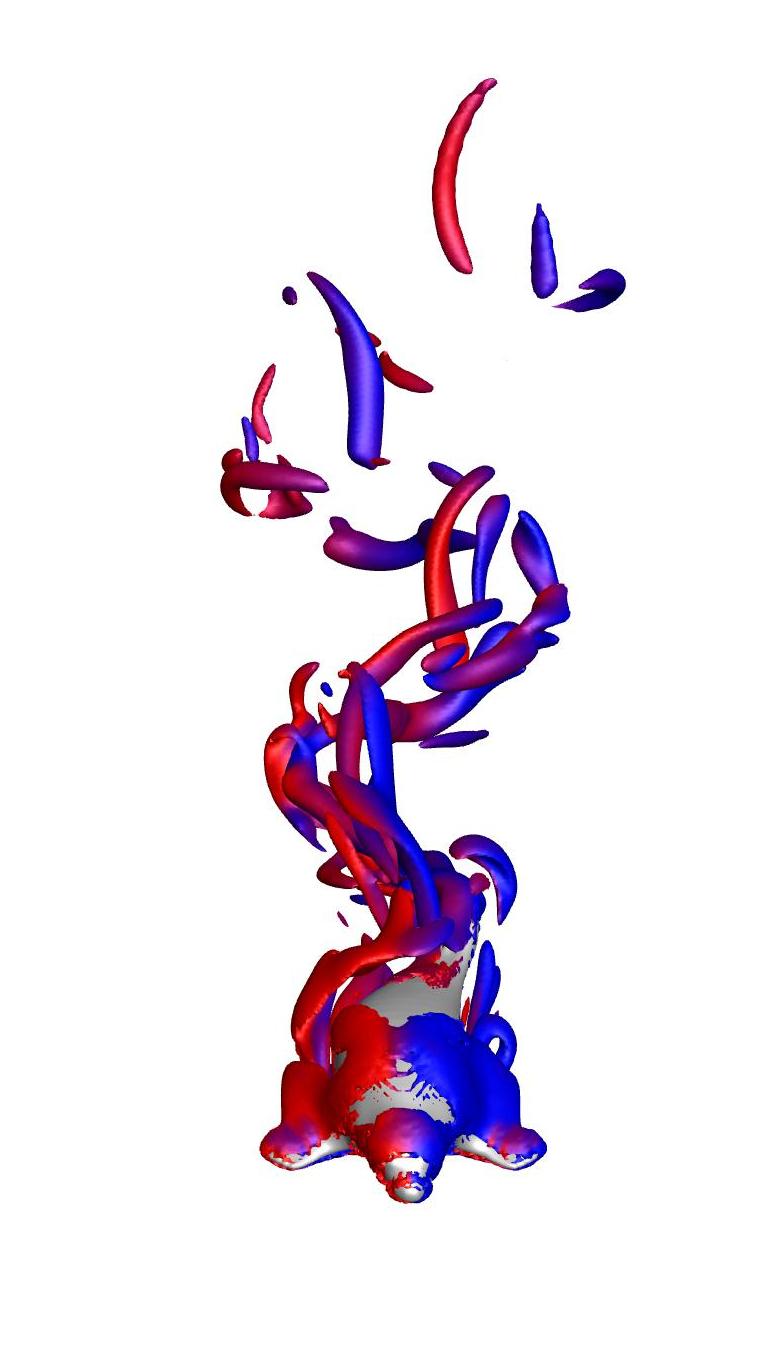}\label{type_St2_1}}
    \end{minipage}
    \begin{minipage}[b]{0.4\linewidth}
        \centering
        \subfloat[(b1)]{\includegraphics[width=0.7\linewidth]{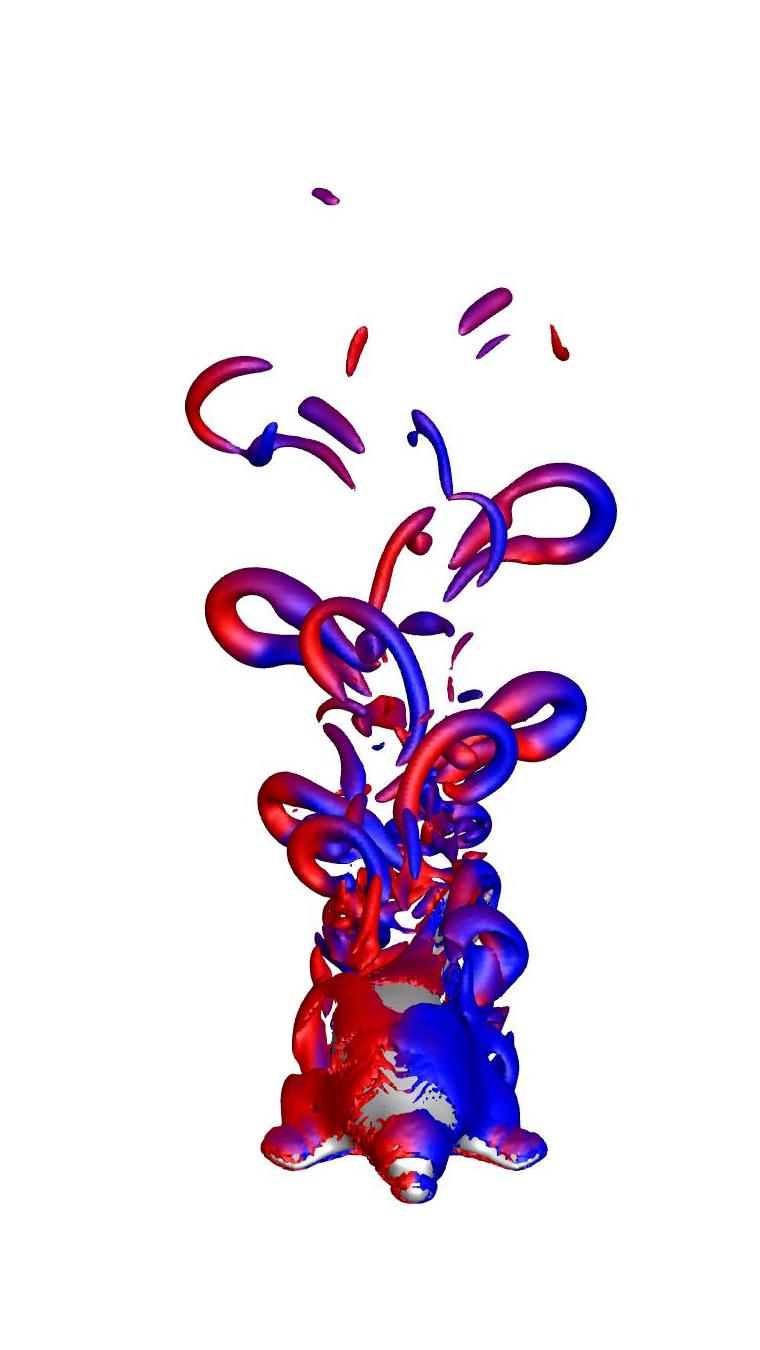}\label{type_St35_1}}
    \end{minipage}
    
    \vspace{0.0cm}
    
    \begin{minipage}[b]{0.4\linewidth}
        \centering
        \subfloat[(a2)]{\includegraphics[width=0.7\linewidth]{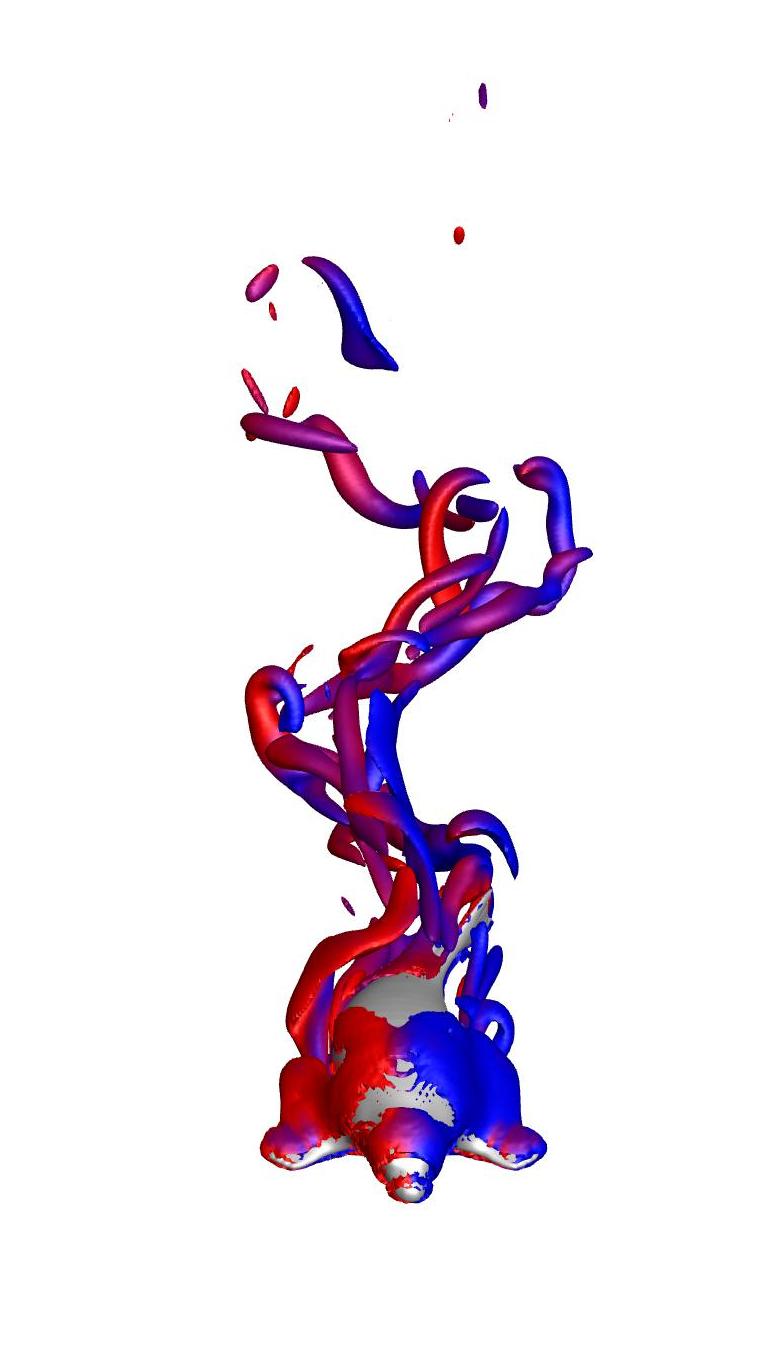}\label{type_St2_2}}
    \end{minipage}
    \begin{minipage}[b]{0.4\linewidth}
        \centering
        \subfloat[(b2)]{\includegraphics[width=0.7\linewidth]{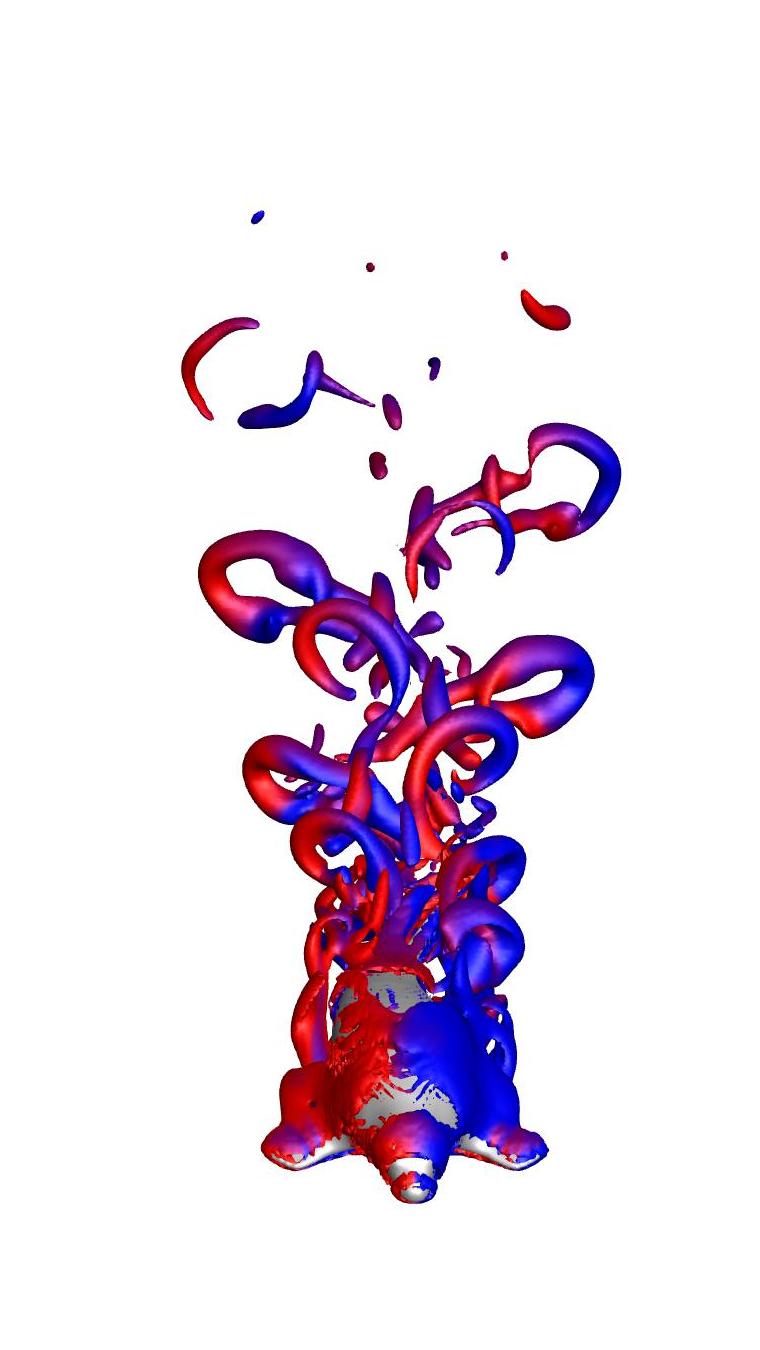}\label{type_St35_2}}
    \end{minipage}
    \vspace{0.0cm}
    \begin{minipage}[b]{0.2\linewidth}
        \centering
        $\omega_y L/U$\\[4pt]
        \subfloat[]{\includegraphics[width=0.7\linewidth]{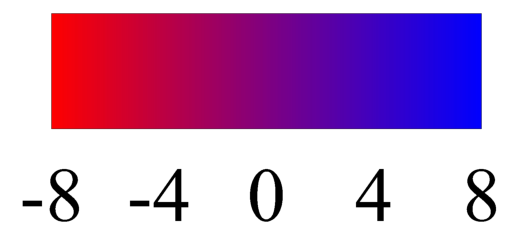}\label{Legend_type}}
    \end{minipage}
    \caption{\small\centering Wake patterns. (a1) and (a2) corresponding to St = 0.2, $\lambda^\ast$ is 1 and 1.2 respectively, (b1) and (b2) corresponding to St = 0.35, $\lambda^\ast$ is 1 and 1.2 respectively }
    \label{Wake_Type}
\end{figure}

\edt{While strongly depending on the key kinematic parameters of $\mbox{St}$ and $\lambda^\ast$,} the wake generated by the \edt{undulating} motions of the harbor \edt{seal} exhibits two distinct patterns, \edt{including a} single-row wake and \edt{a} double-row wake. At lower St, as shown in \edt{Figs.~\ref{Wake_Type}a1 and \ref{Wake_Type}a2}, the wake takes the form of a single-row structure, where vortices remain confined \edt{about} the central axis of the seal's motion. Even increasing the wavelength, the same pattern is observed. However, as $\mbox{St}$ increases \edt{in Figs.~\ref{Wake_Type}b1 and \ref{Wake_Type}b2}, flow transitions to a double-row wake, where vortices are shed in a laterally diverging, wedge-like arrangement. \edt{Similar observations were reported by Borazjani \& Sotiropoulos \cite{borazjani2008numerical} in their numerical study for flows around a} \ahf{mackerel at \fnl{$\mbox{Re} = 4000$}} \edt{that formed single-row and double-row wake streets at lower and higher values of $\mbox{St}$, respectively}. \edt{However, they did not study the influence of undulatory wavelength on the formation of these wake structures and their dynamics}. This \edt{transition} in wake \edt{structures} corresponds to the \ahf{variations in kinematic parameters}  and \edt{the} timing of vortex shedding \edt{from the trailing part of the swimmer}. At \edt{a} higher $\mbox{St}$, increased lateral velocity of the flippers and posterior body induces a stronger lateral flow, driving vortices outward. The resultant double-row wake is distinctly characterized by enhanced three-dimensionality \edt{of the flow} and a wider spatial distribution of vortical structures.

\begin{figure}[htbp]
\centering
{\includegraphics[width=0.8\textwidth]{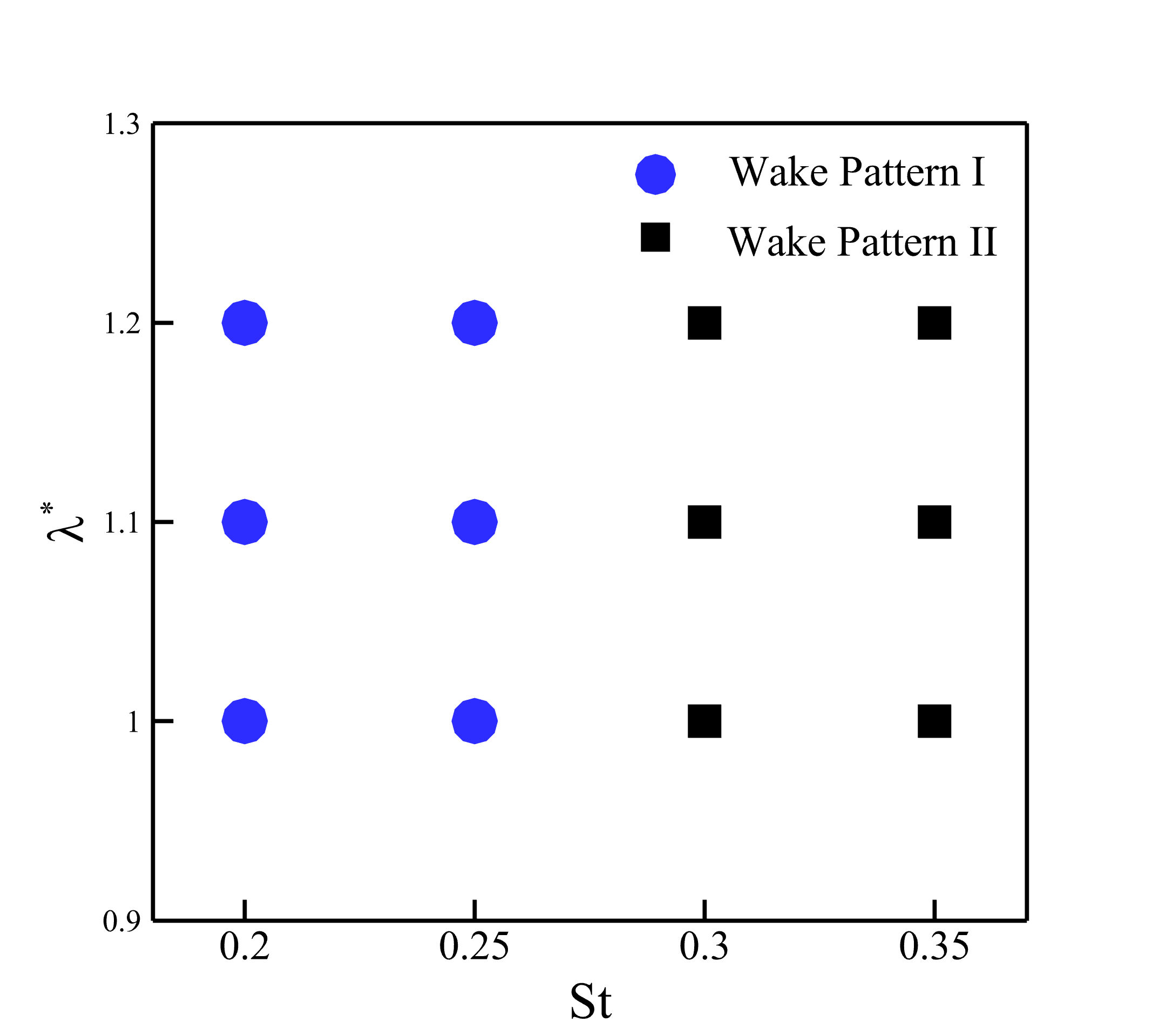}}
\caption{Wake's type classification by St and $\lambda^*$}
\label{Category}
\end{figure}

\edt{Figure}~\ref{Category} \edt{exhibits a phase map for the two specific types of the wake as a function of $\mbox{St}$ and $\lambda^\ast$}. \edt{It is evident here that the type of the wake is primarily dictated by $\mbox{St}$ and not $\lambda^\ast$.}
Based on this classification, we focus on these two representative cases at $\mbox{St}=0.2$ and $0.35$ \ahf{, $\lambda^\ast$ = 1 and 1.2}, which effectively capture distinct wake behaviors \edt{under the flow and kinematic conditions considered in this work}. 

\begin{figure}[h!]
    \captionsetup[subfigure]{labelformat=empty} 
    \centering
    \begin{minipage}[b]{0.4\linewidth}
        \centering
        \subfloat[(a1)]{\includegraphics[width=0.5\linewidth]{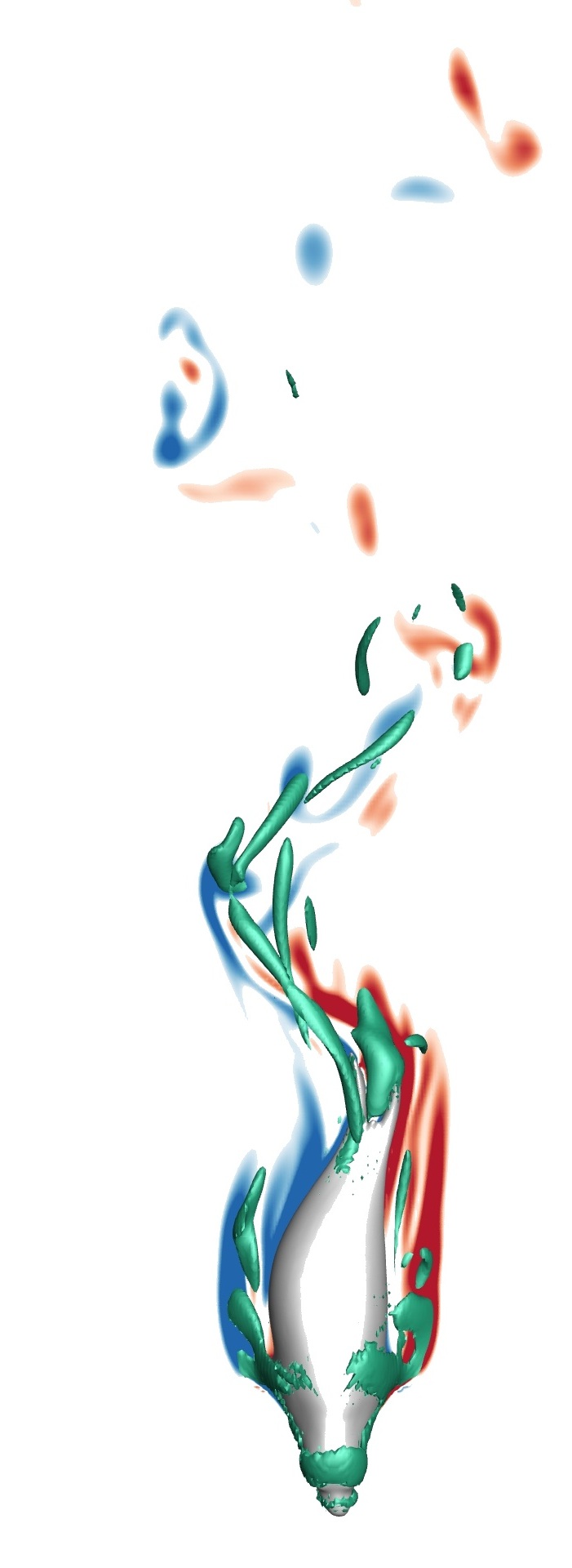}}
    \end{minipage}
    \begin{minipage}[b]{0.4\linewidth}
        \centering
        \subfloat[(b1)]{\includegraphics[width=0.5\linewidth]{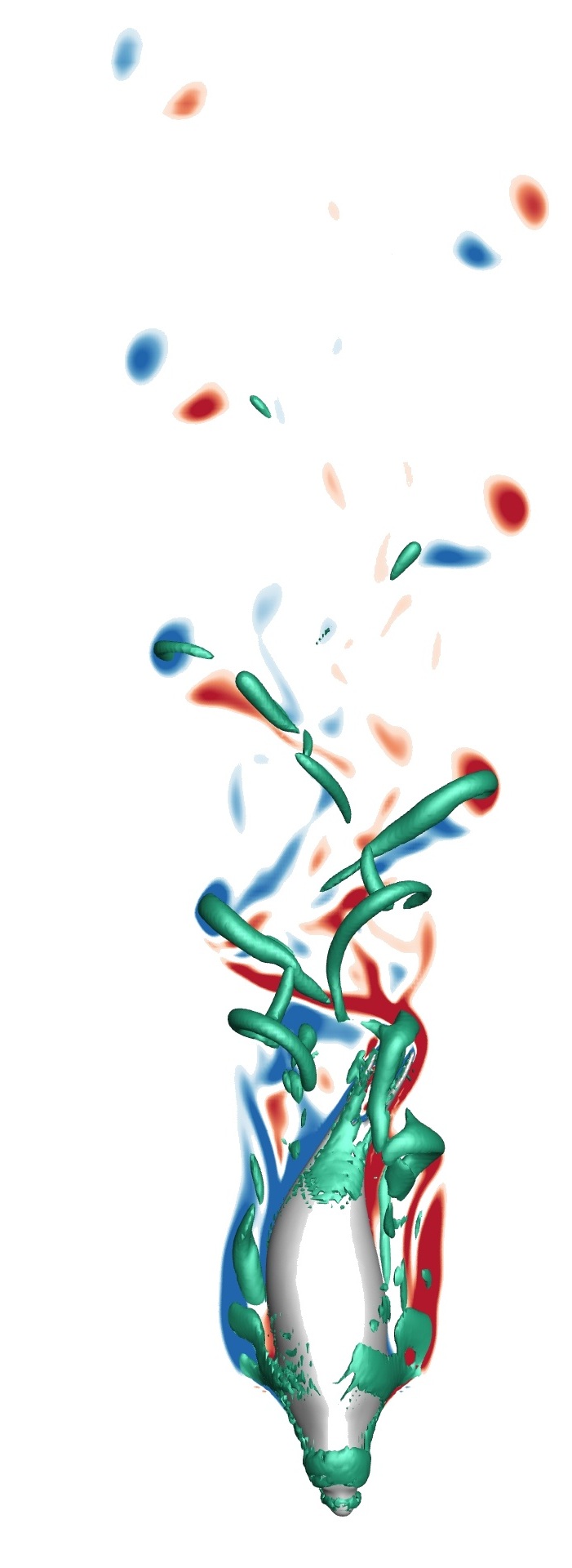}}
    \end{minipage}
    
    \vspace{0.0cm}
    
    \begin{minipage}[b]{0.4\linewidth}
        \centering
        \subfloat[(a2)]{\includegraphics[width=0.5\linewidth]{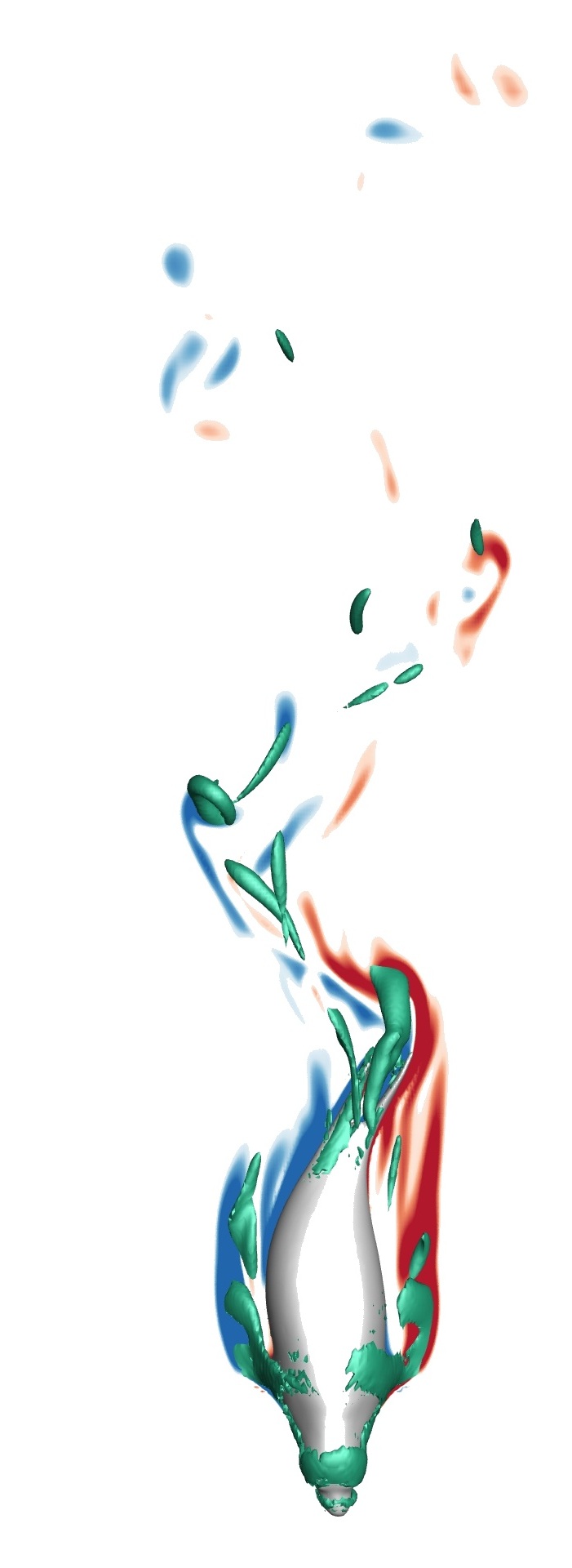}}
    \end{minipage}
    \begin{minipage}[b]{0.4\linewidth}
        \centering
        \subfloat[(b2)]{\includegraphics[width=0.5\linewidth]{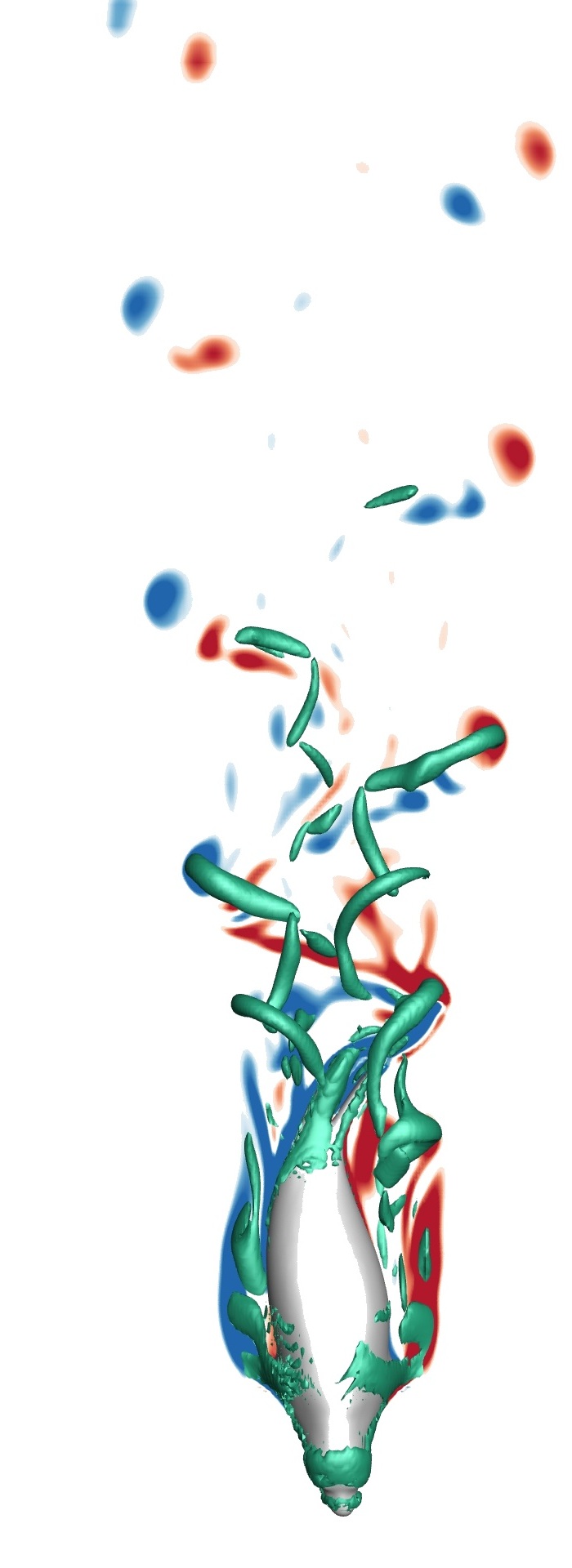}\label{CD2}}
    \end{minipage}
    \vspace{0.0cm}
    \begin{minipage}[b]{0.2\linewidth}
        \centering
        $\omega_y L/U$\\[4pt]
        \subfloat[]{\includegraphics[width=0.7\linewidth]{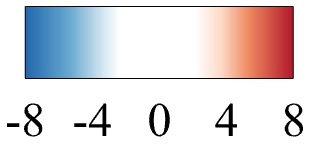}\label{Legend_type}}
    \end{minipage}

    \caption{\small\centering Wake patterns for Strouhal numbers St=0.2 (a) and St=0.35 (b),  configurations at $\lambda^*$ of 1 (a1, b1) and 1.2 (a2, b2).}
    \label{Vertical_View}
\end{figure}

Understanding the topology and dynamics of the three-dimensional coherent structures around the harbor seals and within their wake is crucial, as these structures and their interactions significantly influence hydrodynamic performance. Specifically, we examine the overall wake configurations produced by undulatory motion at varying wavelengths. \ahf{As illustrated in Fig.~\ref{Vertical_View} from the top view}, harbor seals produce and shed multiple vortices of both large and small scales in their wake, highlighting the complex flow characteristics associated with different $\mbox{St}$.

\ahf{A comparison of wavelengths in Fig.~\ref{Vertical_View} reveals that the wake width increases with longer wavelengths. Additionally, \edt{a higher $\mbox{St}$ appears to lead to the formation of more structured and coherent wake patterns. The presence of a core vortex, highlighted in green, is evident in both cases. Notably, at the higher St, vortices are stronger and are shed more prominently into the wake, underscoring the significant impact of St on vortex dynamics and the overall wake structure.}}

\edt{Figure}~\ref{Vortex_Naming}a provides an overview of vortices on \edt{ventral and anterior sides of the body and fore flippers} that form in close proximity to the seal\edt{when the hind flippers undergo the left stroke in their oscillations}. \edt{Here, we identify three vortex structures and shear layers, including ventral body vortex (VBV), fore flipper upper vortex (FFUV), and fore flipper lower vortex (FFLV). VBV is classified as a vortex due to the positive Q-criterion values, which indicate that the rotational component of the flow prevails over the strain rate. It is important to mention that the fore flipper are kept stationary in these simulations, as the scope of our study is limited to investigating the role of hind flippers in determining the hydrodynamics of seals. More importantly, seals are observed to not actively \fnl{utilized} their fore flippers for steady propulsion (see Fig.~\ref{fig:Seal_Video})}. {Structures termed as} FFUV and FFLV rotate in opposite directions, whereas FFUV and VBV \ahf{rotate} in the same direction. These VBVs predominantly interact with FFUV. {As illustrated in Fig.~\ref{Vortex_Naming}b, the hind flipper generates four primary vortices: Left Hind Flipper Upper Vortex (LHFUV), Left Hind Flipper Lower Vortex (LHFLV), Right Hind Flipper Upper Vortex (RHFUV), and Right Hind Flipper Lower Vortex (RHFLV). This figure also highlights the hind flipper vortices formed during the latter half of the stroke. At this specific time instant, Fig.~\ref{Vortex_Naming}b shows that vortices generated by the front flipper and body during the previous half-stroke approach the hind flipper. It is important to note that the vortex generated by the hind flipper plays the most significant role in thrust production.

\begin{figure}[h!]
    \captionsetup[subfigure]{labelformat=empty} 
    \centering
    \begin{minipage}[b]{0.8\linewidth}
        \centering
        \subfloat[(a)]{\includegraphics[width=\linewidth]{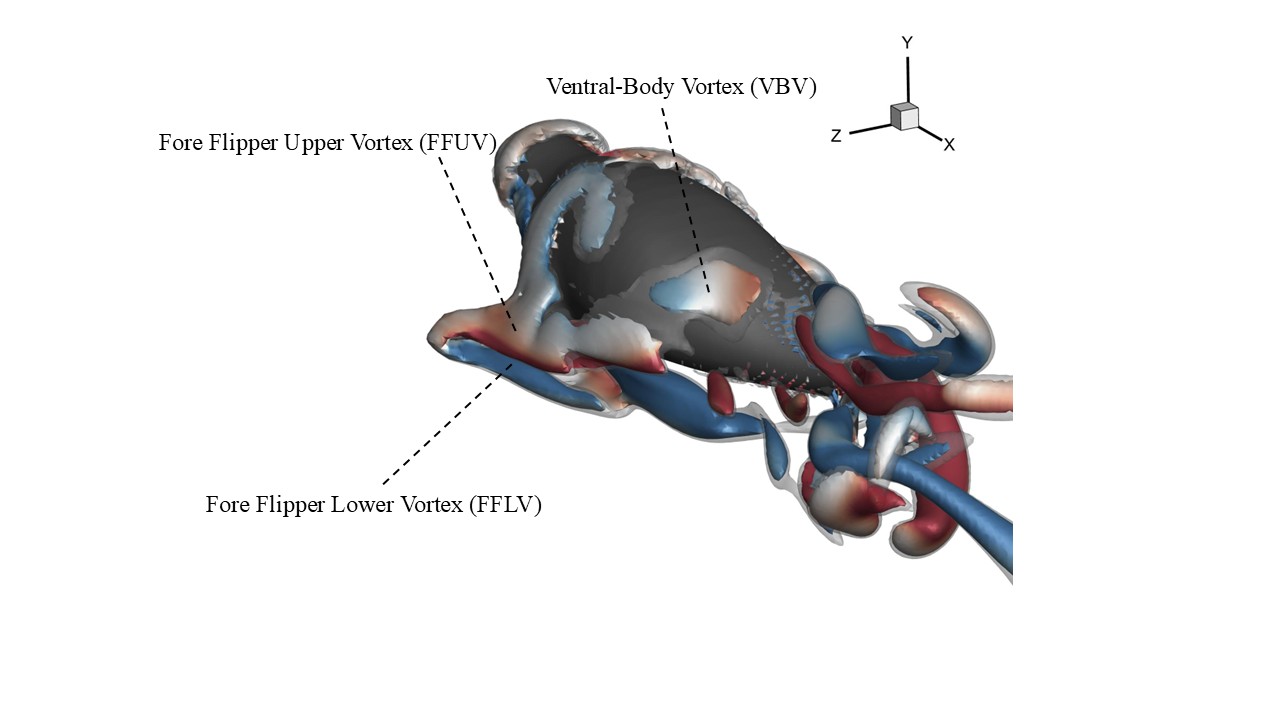}}
    \end{minipage}
    
    \vspace{0.0cm}
    
    \begin{minipage}[b]{0.8\linewidth}
        \centering
        \subfloat[(b)]{\includegraphics[width=\linewidth]{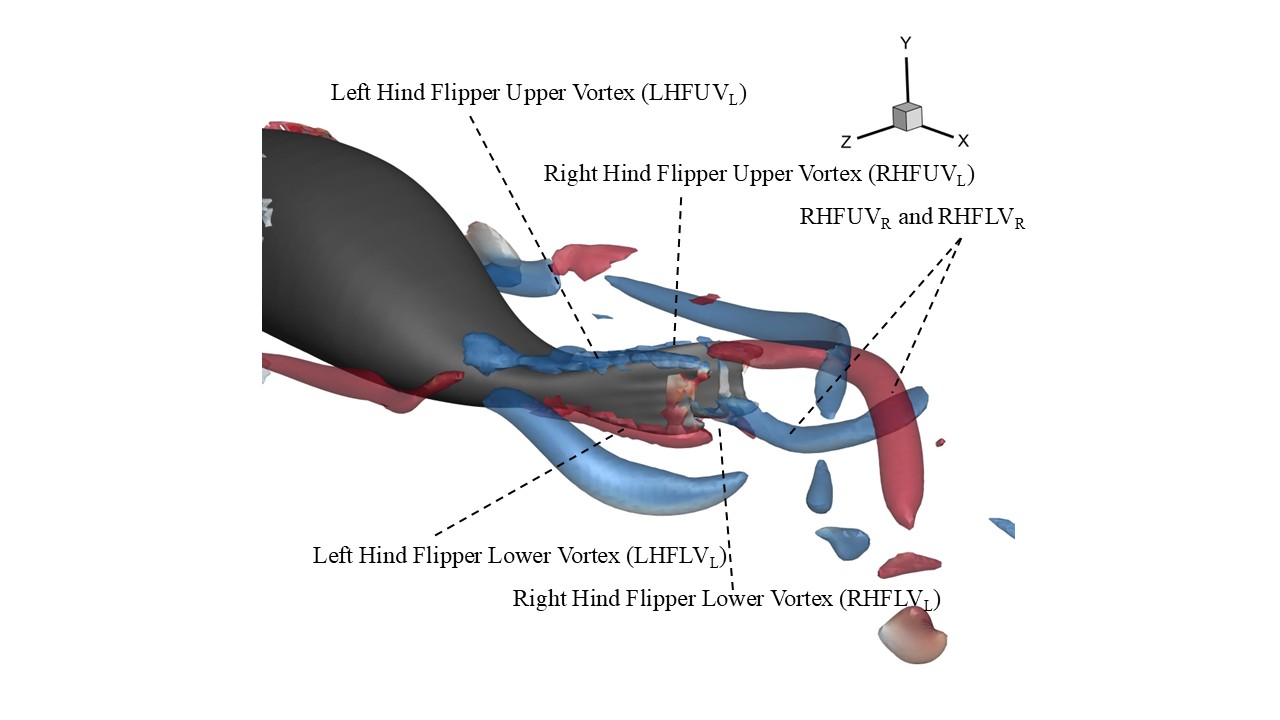}}
    \end{minipage}

    \caption{\small\centering \ahf{Vortex structures around the fore flipper and hind flipper during left stroke}}
    \label{Vortex_Naming}
\end{figure}

\edt{Figure}~\ref{InterferenceST2} \edt{presents} instantaneous vortices at $\lambda^\ast=1.0$ and $1.2$ for $\mbox{St}=0.2$. \edt{Here,} Figs.~\ref{InterferenceST2}a1 -- \ref{InterferenceST2}a5 correspond to $\lambda^* = 1$ \ahf{at five different time instances.} \edt{Figures}~\ref{InterferenceST2}b1 -- \ref{InterferenceST2}b5 \edt{show the vortex dynamics around the swimmer at} $\lambda^\ast=1.2$. \edt{Figure}~\ref{InterferenceST35} presents similar patterns for a Strouhal number of 0.35.
\ahf{The light blue contours represent the Q-criterion at 10, while the blue contours correspond to Q = 50. Please note that we choose different time instants for the comparison of side-by-side frames, because they define the same position of the hind flippers for different $\lambda^\ast$} \ahf{In \edt{F}ig~\ref{InterferenceST2}, at the initial phase (a1 and b1, t/$\tau$= 0.117 and 0.294\edt{, respectively}), we observe the formation of primary vortical structures near the body. VBV develops along the ventral surface while FFUV and FFLV begin to form from the fore-flipper. As \edt{the swimmer advances the undulation,} \fnl{Fig.}\ref{InterferenceST2}a2 and \ref{InterferenceST2}b2, (t/$\tau$ = 0.235 and 0.529, \edt{respectively}), interferences between these vortices become more pronounced. VBV maintains its position while FFUV and FFLV begin to evolve and interact with the surrounding flow field.
In the middle phase \fnl{Fig.}\ref{InterferenceST2}a3 and \ref{InterferenceST2}b3, (t/$\tau$ = 0.705 and 0.764, \edt{respectively}), we see the emergence of \edt{LHFUV$_L$ and LHFLV$_L$.} The key parameter here is the timing of vortex shedding. It is observed that as LHFUV and LHFLV develop, vortex generated by the fore flipper also reaches the hind flipper. The later stages \fnl{Fig.}\ref{InterferenceST2}a4 and \ref{InterferenceST2}b4, (t/$\tau$ = 0.882 and 0.941, \edt{respectively}) show (FFUV) interacts destructively with LHFUV and LHFLV. Additionally, the weak RHFUV and RHFLV can also be observed.
The final phase \fnl{Fig.}\ref{InterferenceST2}a5 and \ref{InterferenceST2}b5, (t/$\tau$ = 0.058 and 0.117, \edt{respectively}) completes the cycle, showing how the vortex structures begin to dissipate and reform, indicating the cyclic nature of the swimming motion. \edt{Also, this} comparison \edt{of flow behavior at} different wavelengths (Figs.~\ref{InterferenceST2}a4 and \ref{InterferenceST2}b4) reveals that increasing the wavelength strengthens the core of the vortices. \edt{We provide a quantitative evidence for this observation} later by \edt{computing} the circulation of vortices. 
}

\ahf{

Figure.~\ref{InterferenceST35} displays a more structured vortex patterns with stronger vortices. At early stages \edt{in the undulatory cycle}, VBV and FFUV, FFLV \edt{begin to} form. \edt{It is clear that} the hind flipper vortex \edt{form} before the fore flippers vortex reach the hind flipper. Here at $\mbox{St}=0.35$, \edt{these vortices do not undergo a destructive interference}. FFUV and Ventral-Body Vortex (VBV) generate stronger vortices that \edt{are} shed separately into the wake. This \edt{phenomenon} features two vortex rings on each side. Compared to $\mbox{St}=0.2$, complete hairpin vortices are formed, indicating a more organized wake structure at the higher Strouhal number. \edt{Besides, a closer examination reveals that} at $\mbox{St}=0.35$, the vortex \edt{attains} a complete hairpin shape, whereas at $\mbox{St}=0.25$, it splits into two distinct vortices.

}




\begin{figure}[htbp]
\centering
{\includegraphics[width=0.65\textwidth]{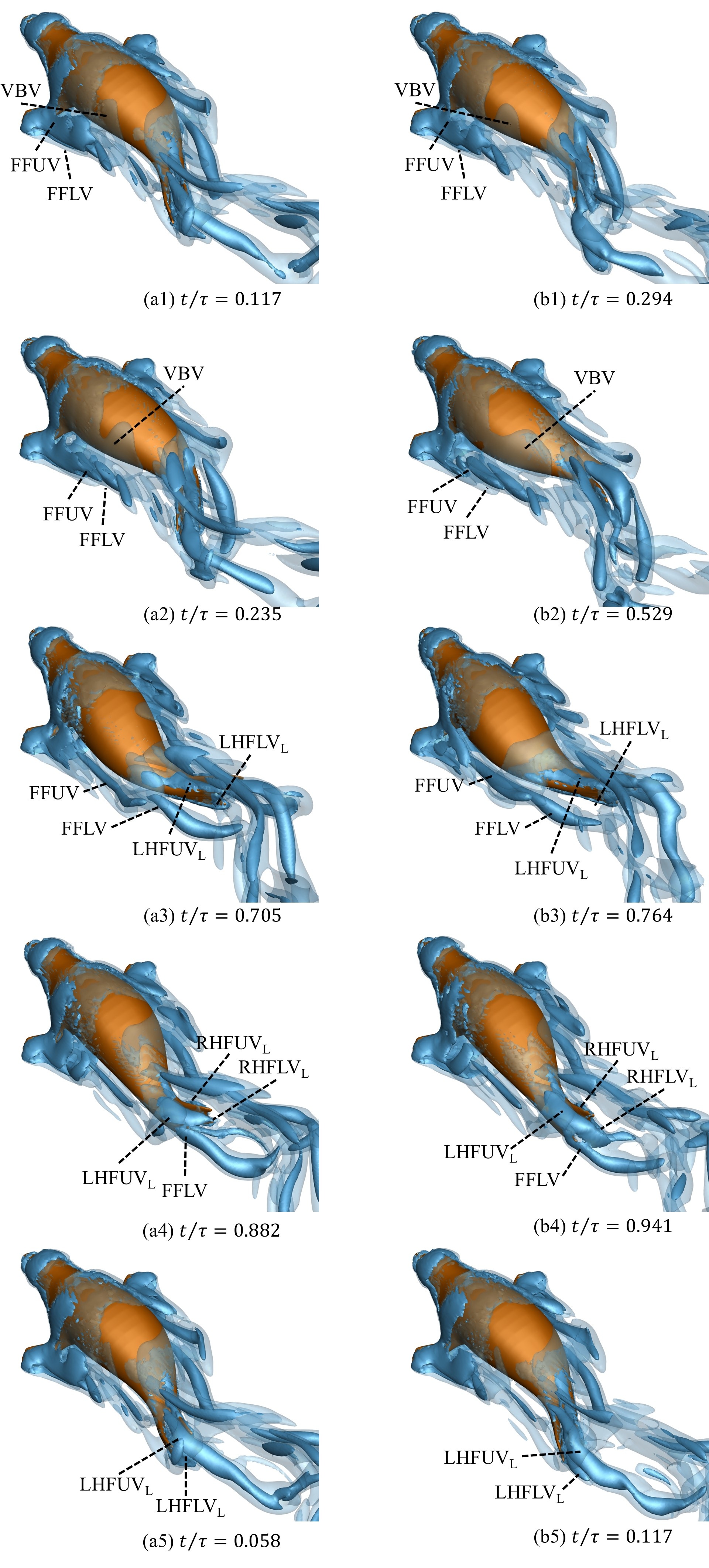}}
\caption{Vortex interferences at St = 0.2.(a) and (b) correspond to $\lambda^*$ = 1 and 1.2, respectively.}
\label{InterferenceST2}
\end{figure}

\begin{figure}[htbp]
\centering
{\includegraphics[width=0.65\textwidth]{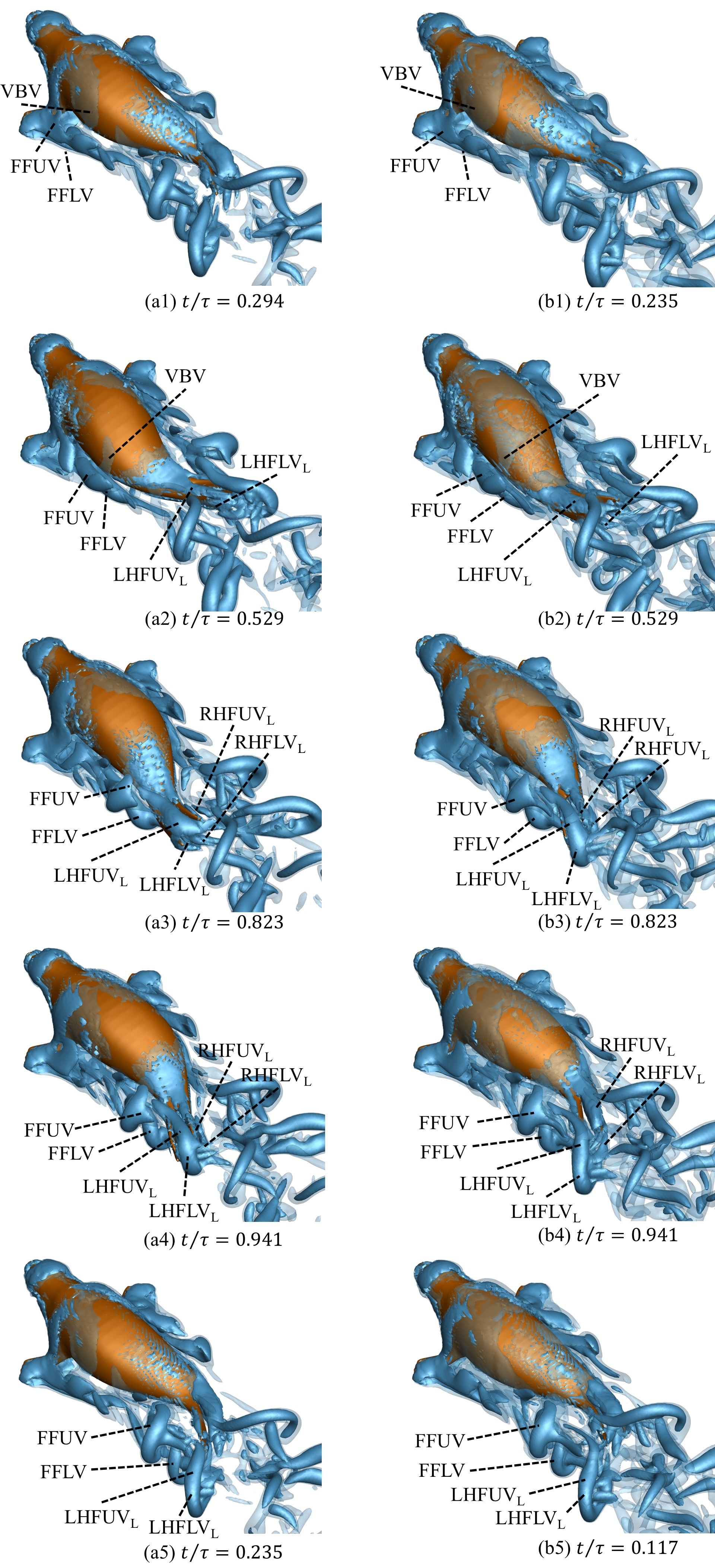}}
\caption{Vortex interferences at St = 0.35.(a) and (b) correspond to $\lambda^*$ = 1 and 1.2, respectively.}
\label{InterferenceST35}
\end{figure}

Figure~\ref{Left_stroke} illustrates the streamwise \edt{vortices} at \edt{a section located at }$95\%$ of the body length. Analyzing panels Figs.\ref{Left_stroke}a and \ref{Left_stroke}b, which correspond to $\mbox{St}=0.2$, reveals destructive interference between a vortex generated by the front flipper and that from the hind flipper. At the same $\mbox{St}$, increasing $\lambda^\ast$ reduces this destructive interaction as vortices become more spatially separated.

In contrast, examining panels Figs.\ref{Left_stroke}c and \ref{Left_stroke}d, which represent $\mbox{St}=0.35$, shows no interference between these vortices due to the altered timing of vortex shedding. This absence of interference allows for greater thrust production \edt{because vortices grow in their sizes around the hind flippers in these cases and this increase in size leads to greater pressure differences, as discussed by Khalid et al.\cite{khalid2021larger}}. Furthermore, as $\mbox{St}$ increases, strength of vortices \edt{is enhanced}, which is \edt{also} reflected in the \edt{circulation of these vortices presented next}. In the left stroke, as shown here, majority of thrust is produced by the Left Hind Flipper Vortex. While the Right Hind Flipper also contributes to thrust, its magnitude is significantly smaller compared to \edt{that on the} left flipper. \edt{It indicates} the asymmetrical contribution of the flippers to propulsion during the left and right strokes.

\begin{figure}[htbp]
\centering
{\includegraphics[width=0.8\textwidth]{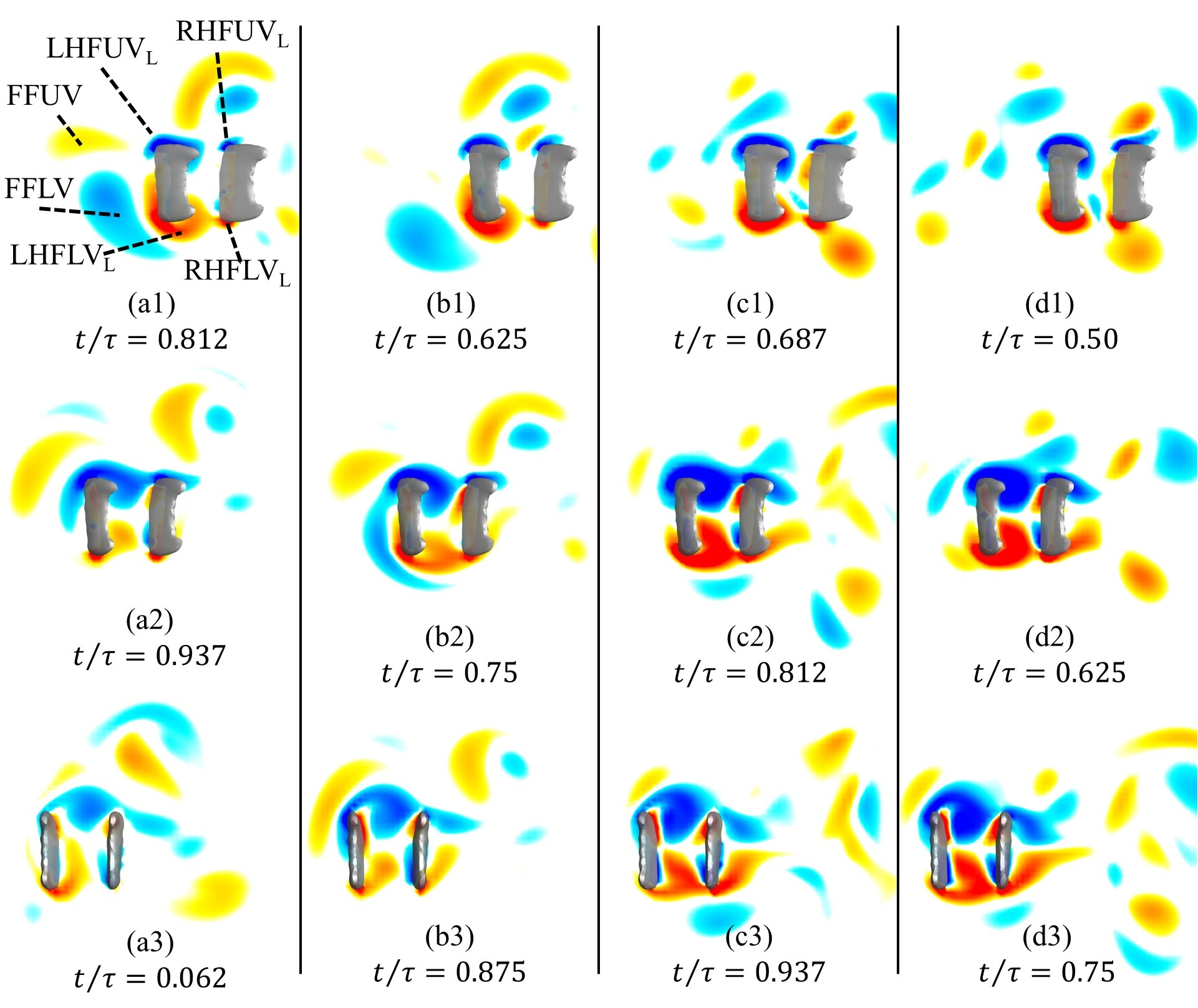}}
\vspace{0.0cm}
    \captionsetup[subfigure]{labelformat=empty}
    \begin{minipage}[b]{0.2\linewidth}
        \centering
        $\omega_xL/U$\\[4pt]
        \subfloat[]{\includegraphics[width=0.7\linewidth]{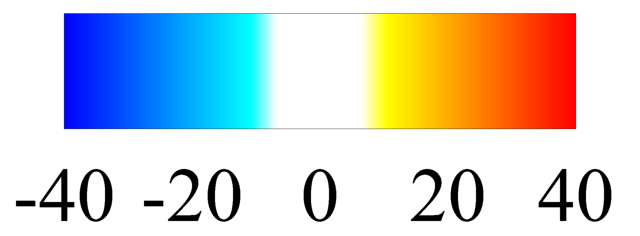}}
    \end{minipage}
\caption{Streamwise vorticity component at 95\% of the body length, captured at the start, middle, and end of the left stroke. Panels (a) and (b) correspond to St=0.2 with $\lambda^* = 1$ and $\lambda^* = 1.2$, respectively, while panels (c) and (d) represent St=0.35 with $\lambda^* = 1$ and $\lambda^* = 1.2$, respectively.}
\label{Left_stroke}
\end{figure}

In addition to this qualitative visualization, we compute the circulation ($\Gamma$) of vortices \edt{located at} $95\%$ of the body during the left stroke oscillation cycle for each case shown in Fig.~\ref{Left_stroke}. For this purpose, we employ the methodology previously \edt{developed} and reported by \edt{Khalid et al.} \cite{khalid2020flow}. Circulation is a measure of the strength of a vortex and is mathematically defined either as the line integral of the velocity field over its boundary:
\[
\Gamma = \oint_{} \mathbf{V} \cdot d\mathbf{l},
\]
\noindent or as the surface integral of the vorticity field over the area of the vortex:
\[
\gamma = \int\int_{S} \omega \cdot dS.
\]

To ensure accurate computation and to avoid overlap of a vortex boundary with neighboring vortices \cite{godoy2009model}, \edt{the devised technique} allows the manual prescription of four points around a vortex through visual inspection. Subsequently, the \edt{algorithm automatically} identifies all data points with positive or negative vorticity greater or smaller than a predefined threshold value. In this study, we define the threshold as $1\%$ of the maximum value of vorticity ($\omega_x$) in the flow field. This approach ensures the extraction of a single vortex without interference from surrounding vortices. Once the vortex is identified, we draw a boundary encompassing it without imposing any predefined geometric shape. Finally, we perform numerical integration of the dot product of the velocity components and their respective displacement vectors, yielding accurate and reliable measurements of the vortex strength.

\edt{We present the values of nondimensional circulation for over one complete undulation cycle for different Strouhal numbers and wavelengths.} As shown in Fig.~\ref{Circulation}, circulation of \edt{the vortex around the} left flipper is stronger than that of the right flipper. It is important to note that circulation values in Figs.~\ref{Circulation}a and \ref{Circulation}b are negative due to the axis of rotation, indicating the presence of negative vortices. However, magnitude of the circulation is significant in this context. This quantitative analysis \edt{elucidates} the influence of two governing kinematic parameters ($\mbox{St}$ and $\lambda^\ast$) on thrust production, showing that increasing either parameter results in a significantly increased thrust.

Also from Fig.~\ref{Circulation}, we observe destructive interference between vortices generated by the fore flipper and the hind flipper. Specifically, Fig.\ref{Circulation}c highlights that at the initial stage, the vortex strength for $\mbox{St}=0.35$ is lower than that for $\mbox{St}=0.2$ and $\mbox{St}=0.25$. However, destructive interference diminishes the vortex strength at lower $\mbox{St}$ values, whereas for higher $\mbox{St}$, the strength of the vortex continues to increase.

\begin{figure}[htbp]
\centering
{\includegraphics[width=0.8\textwidth]{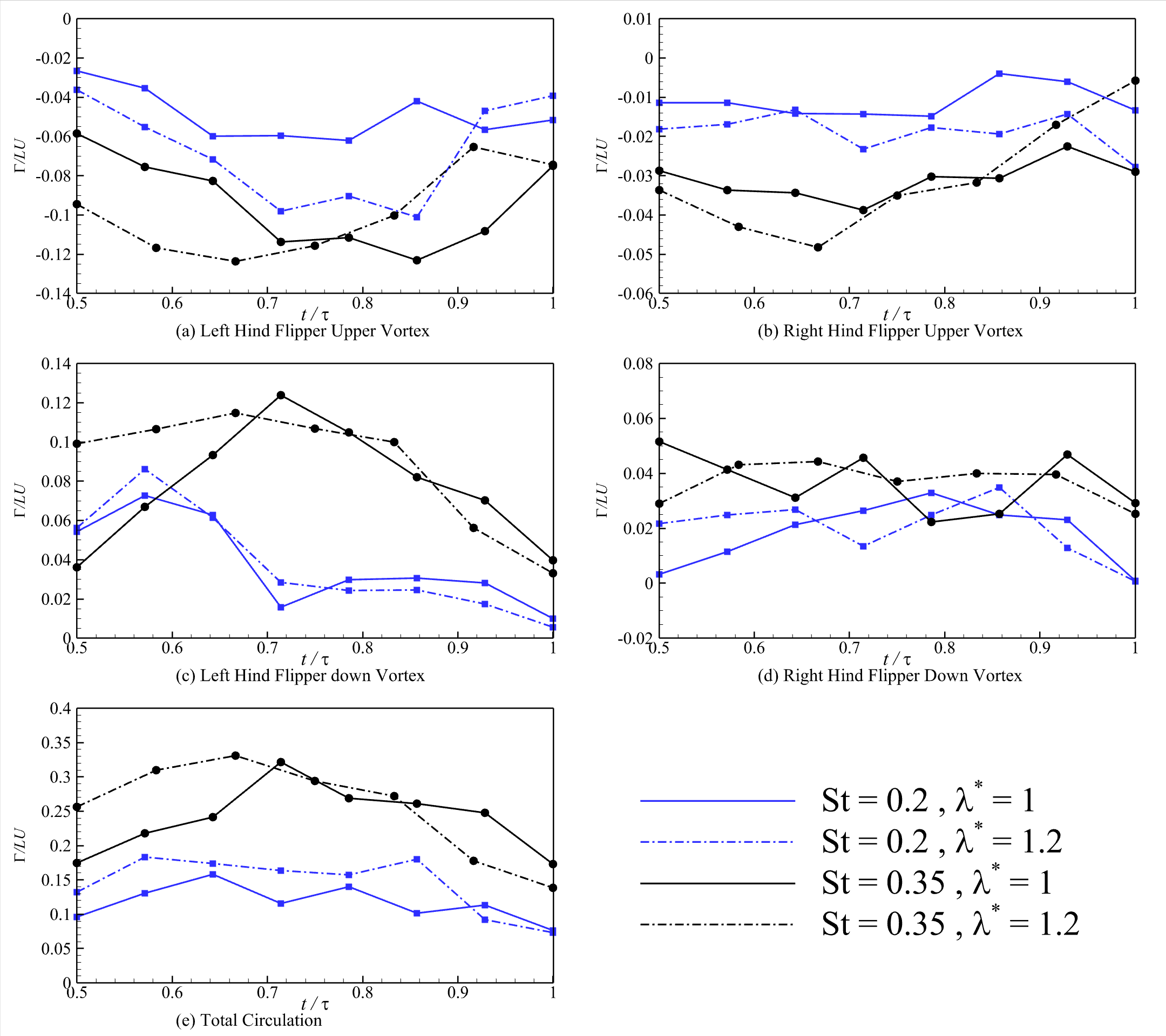}}
\caption{Circulation of the streamwise vortex at 95\% of the body length for each hind flipper during the left stroke (panels a, b, c, d) along with the total circulation value (e).}
\label{Circulation}
\end{figure}


\section{Conclusions}
Our comprehensive analysis of swimming dynamics of a harbor seal reveals several key findings that advance our understanding of aquatic locomotion. First, the study establishes that swimming \edt{performance} is significantly influenced by both Strouhal number and wavelength, with higher values of both parameters generally leads to improved thrust generation. Transition from single-row to double-row wake patterns at $\mbox{St}=0.3$ marks a crucial shift in propulsion mechanics, with the latter configuration associated with more structured and efficient vortex formations. Investigation of vortex interactions between front and hind flippers demonstrates that timing is crucial for optimal propulsion. At lower Strouhal numbers ($\mbox{St}=0.2$), destructive interference between the front flipper and hind flipper vortices reduces thrust. \edt{Contrarily}, at higher Strouhal numbers ($\mbox{St}=0.35$), the altered timing of vortex shedding prevents this interference, resulting in more effective thrust production. Analyzing circulation quantitatively confirms that increasing both Strouhal number and wavelength strengthens vortex formation, particularly during the left stroke. This finding suggests that harbor seals might preferentially operate at higher wavelengths and Strouhal numbers for optimal swimming performance, contrary to some other marine species like anguilliform swimmers. These results not only enhance our understanding of seal's locomotion but also provide valuable insights for the development of bio-inspired propulsion systems. Future research could explore the application of these findings in underwater vehicle design, particularly in developing more efficient propulsion mechanisms based on the dual-flipper system observed in seals.

\section*{Acknowledgement}
MSU Khalid acknowledges the funding support from the Natural Sciences and Engineering Research Council of Canada (NSERC) through the Discovery grant program for this work. The simulations reported in this work were performed on the supercomputing clusters administered and managed by the Digital Research Alliance of Canada. 

\vskip6pt









\bibliography{Draft_Manuscript}



\end{document}